\documentclass[conference]{IEEEtran}
\IEEEoverridecommandlockouts
\usepackage{cite}
\usepackage{amsmath,amssymb,amsfonts}
\usepackage{algorithmic}
\usepackage{graphicx}
\usepackage{amsmath}
\usepackage{array}
\usepackage{makecell}
\usepackage{longtable}
\usepackage{graphicx}
\usepackage{lipsum}
\usepackage{float}
\usepackage{multicol}
\usepackage{multirow}
\usepackage{textcomp}
\usepackage{xcolor}
\def\BibTeX{{\rm B\kern-.05em{\sc i\kern-.025em b}\kern-.08em
    T\kern-.1667em\lower.7ex\hbox{E}\kern-.125emX}}
\begin{document}

\title{Cyber-Attack Event Analysis for EV Charging Stations\\
}

\author{\IEEEauthorblockN{Mansi Girdhar and Junho Hong}
\IEEEauthorblockA{\textit{Electrical and Computer Engineering} \\
\textit{University of Michigan-Dearborn}\\
Dearborn, USA \\
gmansi@umich.edu \\
jhwr@umich.edu}
\and
\IEEEauthorblockN{Yongsik You and Tai-jin Song}
\IEEEauthorblockA{\textit{Department of Urban Engineering} \\
\textit{Chungbuk National University}\\
Cheongju, South Korea \\
yys@chungbuk.ac.kr \\
tj@chungbuk.ac.kr}
\and
\IEEEauthorblockN{Manimaran Govindarasu}
\IEEEauthorblockA{\textit{Electrical and Computer Engineering} \\
\textit{Iowa State University}\\
Ames, USA \\
gmani@iastate.edu}
}

\maketitle

\begin{abstract} 
Safe and secure electric vehicle charging stations (EVCSs) are important in smart transportation infrastructure. The prevalence of EVCSs has rapidly increased over time in response to the rising demand for EV charging. However, developments in information and communication technologies (ICT) have made the cyber-physical system (CPS) of EVCSs susceptible to cyber-attacks, which might destabilize the infrastructure of the electric grid as well as the environment for charging. This study suggests a 5Ws \& 1H-based investigation approach to deal with cyber-attack-related incidents due to the incapacity of the current investigation frameworks to comprehend and handle these mishaps. Also, a stochastic anomaly detection system (ADS) is proposed to identify the anomalies, abnormal activities, and unusual operations of the station entities as a post cyber event analysis. 

\end{abstract}


\section{INTRODUCTION}
EVs, which are perceived as a solution to climate change, are replacing internal combustion engines as the dominant form of transportation in the global auto industry \cite{9215386}. Because of the EVs' steady expansion, EVCSs are now more prevalent in electric grids. Since the market value of EVCSs is estimated to exceed \$103.6 billion by 2028, it is critical to have a well-thought-out plan for deploying EVCSs and sturdy supporting electric distribution infrastructures. As a result, engineers and researchers specializing in the power and transportation systems have recently performed an extensive study to promote the development and diffusion of this game-changing technology. Also, a number of industry sectors and original equipment manufacturers (OEMs), including General Motors (GM), Tesla, ABB, Atom Power, BTCPower, ChargePoint, EVBox, and EV Connect, are investing billions in the research and development of EVCSs \cite{9786788}. 

In general, EVCS is a safety-critical system composed of many heterogeneous components, both physical and cyber, that could provide significant security concerns. It often functions as a high-wattage gateway or bridge that supports the energy exchange between an EV and the power grid, thus facilitating EV charging services \cite{9336258}. Cyber-attack methods on critical infrastructures, particularly energy delivery systems, are maturing and becoming more sophisticated, owing to the internet-of-things (IoT) paradigm, which has created multiple vulnerabilities. Since internet-connected EVCS is a key entity in an upcoming smart grid, the cybersecurity stance of EVCS is crucial to the functioning of power systems. For instance, the shared data may be susceptible to various vulnerabilities because of the intricate cyber-physical interactions and interdependencies existing at the confluence of EVs, EVCSs, and the power grid. Also, the adversaries might exploit these known vulnerabilities of the EVCS ecosystem to breach the system, thus endangering its security. Hence, it is crucial to analyze the bi-directional interactions between the distribution grid, EVs, and charging infrastructure \cite{9272723}. Therefore, white-hat hackers and researchers are very interested in addressing the cybersecurity aspects of EVCSs due to the serious consequences that may be caused by hostile cyber-attacks.

Even though there has been a significant amount of research on designing successful and robust cybersecurity frameworks for electric grid domains, e.g., the critical infrastructure protection (CIP) 002-009 standard by the North American Electric Reliability Corporation (NERC), the IEC 62351 standard \cite{9369743}, and ADSs \cite{7785783}, well-trained attackers could exploit the vulnerabilities of EVCSs \cite{9916758}. Similarly, ISO/IEC 15118 standardizes wireless vehicle-to-grid (V2G) communication \cite{8493213}. However, these cybersecurity measures do not improve the security of EVCSs.

Although, there have not yet been any notable or well-publicized cyber events with EVCSs, scientists (e.g., Kaspersky Lab researchers) have been meddling with charging operations to see whether any undiscovered vulnerabilities exist \cite{905746}. Therefore, some rigorous work has been done to address the key cybersecurity gaps in an EVCS, e.g., cybersecurity recommendations from the US DOE, NHTSA, and DOHS. However, these proposed solutions have not yet been formalized and put into effect \cite{7807218}.

Additionally, due to the presence of different entities, it is uncertain how an EVCS can adapt to various scenarios and settings in several cybersecurity risks that could interfere with its operation. Thereby, cyber-induced EVCS event investigation becomes highly complicated and challenging in such a complex environment.

The underlying contributions of this paper are: (1) the development of a post-cyber event investigation framework based on 5Ws \& 1H model, and (2) the proposal of a probability-based anomaly detection algorithm in an EVCS cyber-induced incident, along with a case study. 

The remainder of the paper is divided as follows:\\
Section II discusses the high-level overview of an EVCS environment. Sections III analyzes the cyber-attack induced events caused in the station environment. Further, a digital forensic investigation framework based on 5Ws \& 1H to find the causes and effects of an incident is outlined in section IV. Section V proposes an ADS to identify the abnormal behavior of the station network entities. Section VI analyzes the performance of the proposed frameworks by engaging a case study of cyber-attacks on EVCS. Finally, section VII concludes the paper along with the limitations and recommendations for future work.

\section{HIGH-LEVEL OVERVIEW OF AN EVCS ENVIRONMENT}
There have been several digital architectures proposed in the past to model the environment of an EVCS. The work of \cite{9583592} introduces an EVCS model, which consists mainly of EV chargers (EVCs), grid components (e.g., step-down distribution transformer, circuit breakers (CBs)), battery energy storage system (BESS), battery management system (BMS), cloud system and open charge point protocol (OCPP). In another work, \cite{9490069}, the authors present an architecture that identifies legacy grid components, vendor cloud server, EVs, charging network operators, and different communication standards (e.g., IEC 62196-1/2/3, SAE J1772, ISO 15118, ISO 61851-24, ISO 61850, and IEEE 2030.5), as distinct entities comprising the EVCS environment. It thereby correlates the security of an EVCS ecosystem to the security of every subsystem. In other words, any inaccurate cyber interactions may result in an unusual event that significantly influences the station environment. Hence, the complex mechanism of a malicious EVCS event may be caused due to multiple endogenous or exogenous trigger conditions, leading to a hazardous condition in one or more interconnected elements present in its ecosystem. For instance, attackers may use the hacked EVCS to manipulate the charging behaviors of the station in a gradual way, endangering the supply-demand balance of the grid during peak hours. ALso, attackers may produce an extended period of high demand, which could lead to a cascade disconnect of the supply from the power grid and aberrant functioning. The results would be multiple outages, blackouts, or system instability due to the instability of the power system operations. Hence, it can threaten the security and stable operation of the power systems. Similarly, the attacker may get access to the EVCS through the compromised EVs, which could have cascade repercussions \cite{8030362}. 

\section{CYBER-ATTACK EVENT IDENTIFICATION}
\subsection{Cyber-Attack Induced EVCS Incident Analysis}
A radical transformation in the hyper-connected EVCS environment brought new cybersecurity concerns. For instance, existing EVCSs use big data, the cloud, V2G communication technology, and internal communication to exchange information. Poorly managed ICT systems are susceptible to various attacks, including attacks on hardware (e.g., electronic control devices and sensors) and software (e.g., malware attacks on firmware) \cite{9698889}.

\begin{table*}[t]
\centering
\caption{Proposed 5Ws \& 1H-based digital forensic investigation of a EVCS cyber incident.}
\label{tab:investigation}
\begin{tabular}{|l|l|l|}
\hline
\multicolumn{1}{|c|}{\textbf{Attribute}} &
  \multicolumn{1}{c|}{\textbf{Definition}} &
  \multicolumn{1}{c|}{\textbf{Dataset}} \\ \hline
\multirow{2}{*}{Who} &
  Attacker ($\alpha$) &
  $\alpha$ = \{$\alpha$1: Hacker, $\alpha$2: spy, $\alpha$3: terrorist, $\alpha$4: vandal, $\alpha$5: raider\} \\ \cline{2-3} 
    & Victim ($\nu$)  & \makecell[l]{$\nu$ = \{$\nu$1: EV, $\nu$2: power grid, $\nu$3: cloud system, $\nu$4: communication protocols\}}    \\ \hline
    
What &
  Target ($\tau$) &
  $\tau$ = \{$\tau$1: OCPP, $\tau$2: BMS, $\tau$3: charging adapter, $\tau$4: cooling system, $\tau$5: smart meter, $\tau$6: HMI\} \\ \hline
\multirow{2}{*}{When} &
  Date ($\delta$) &
 $\delta$ = \{$\delta$1: Month, $\delta$2: date, $\delta$3: year\},
 Format: (mm-dd-yy) \\ \cline{2-3} 
 &
  Time ($\iota$) &
  $\iota$ = \{$\iota$1: Timezone , $\iota$2: hours, $\iota$3: minutes, $\iota$4: seconds, $\iota$5: milliseconds\}, Format: (hh:mm:sec:msec) \\ \hline
Where &
  Attack path ($\rho$) &
 $\rho$ = \{$\rho$1: OTA, $\rho$2: software kickout,  $\rho$3: incorrect coding\} \\ \hline
Why &
  Hazardous behavior ($\beta$) &
  $\beta$ = \{$\beta$1: Faulty SOC, $\beta$2: unintended overcharging, $\beta$3: incorrect scheduling, $\beta$4: system malfunction\} \\ \hline
How & Attack method ($\omega$) & $\omega$ = \{$\omega$1: Spoofing, $\omega$2: tampering,  $\omega$3:  repudiation, $\omega$4: information disclosure, $\omega$5: denial of service\} \\ \hline

\end{tabular}
\end{table*}
 
The damage caused by risks or hazards could be managed by appropriate mitigation procedures. In the worst case, the mitigation actions will be failed, and the EVCS cannot provide a normal operation. However, the EVCS could revert to a healthy state if the system engineers implement appropriate physical (e.g., security guards and surveillance cameras) and technical (e.g., firewalls and multifactor authentication) control measures to identify, mitigate, and prevent the cyber threats and attacks. Further, controllability can also be provided by designing intrusion detection and prevention (ID\&P) systems to avert potential system failures. Therefore, this paper proposes an incident-based post-event analysis model by tracking the involvement of a particular candidate.

Furthermore, an EVCS incident in a multi-model environment can be mathematically formulated as,
\begin{equation}
\sum_{i=1}^{n}\sum_{j=1}^{n}(\zeta_{i}\times \chi_{j})\ge 1, 
\end{equation}
where,
\begin{equation}
\zeta_{i} = \left\{ \begin{array}{cl}
1 & : \ \text{entity}(\mathrm{E}_{i}^{k})\text{ } \text{suffers a cyber-attack} \\
0 & : \ \text{no cyber-attack}
\end{array} \right.
\end{equation}

\begin{equation}
\chi_{j} = \left\{ \begin{array}{cl}
1 & : \ \text{no controllability by mitigating actors}(\mathrm{M}_{j}^{l})\\
0 & : \ \text{successful controllability}
\end{array} \right.
\end{equation}
\\
The symbols used in the above equations are defined as,
\\
E$^{\text{k}}_{\text{i}}$: ith component of the k entity or subsystem in the EVCS environment, where k = \{G, V, O, P, S\}
\\
G: Power grid
\\ 
V: EV
\\
O: Cloud server
\\
P: Communication protocols
\\
S: Station infrastructure
\\
M$^{\text{l}}_{\text{j}}$: jth component of the mitigating features responsible for averting the incident, where l = \{C, I\}
\\
C: Physical and technical controls
\\
I: ID\&P
\section{DIGITAL FORENSIC INVESTIGATION FRAMEWORK}
Lack of cybersecurity measures and forensics capability in the EVCS may lead to serious consequences if the vulnerability is effectively exploited and a cyber-attack is successfully conducted. The internet-enabled EVCS and the associated infrastructure need to have effective protection against potential cyber-attacks. Trustworthy computing methods are required to ensure the reliability of the applications used and the data gathered in the charging station environment. There have been a few systematic forensic investigations carried out in the literature in the areas of smart power grid \cite{905fg578}, \cite{KORONIOTIS202091}, and \cite{9243536} and connected and automated vehicles (CAVs) \cite{9849671}. However, to the best of the authors' knowledge, there is no forensic investigation framework designed specifically for EVCSs. Therefore, this section describes the development of the proposed EVCS cyber-attack event analysis framework for digital forensics investigations. 

The framework consists of several processes, e.g., log integration/acquisition, preprocessing, correlation, sequencing, analysis, and reporting. When an event occurred at the EVCS, the investigative team receives official approval after receiving requests to conduct an inquiry from internal or external parties. It is followed by log gathering and extraction from all the EVCS devices. Additionally, data preprocessing is carried out to undertake data cleaning and filtering with the goal of focusing only on important data in the log. The subsequent stages involve correlation and sequencing to create a chronology based on the time of the events. Then, the investigation starts with gathering evidence (e.g., imaging data and asset seizure), followed by evidence analysis using forensic tools and techniques, identifying the 5Ws \& 1H, and presenting the findings. Finally, the crime scene is documented by being recorded, photographed, sketched, mapped, and recorded.

 The incident responders can use the 5Ws \& 1H to prepare, summarize, and report their discoveries in an investigation that includes the six key questions, as depicted in Table~\ref{tab:investigation}, in order to appropriately document and analyze the cyber incident scene. First, it exhibits 5Ws \& 1H, defined as, (1) \textbf{W}ho, which identifies the kind of attacker and the station environment entity that is being attacked, (2) \textbf{W}hat, specifying the system failure or attack target, (3) \textbf{W}hen, describing the date and time of the cyber incident and the failure's occurrence, (4) \textbf {W}here, providing the location of the cyber event or the route taken by the attacker for each EVCS function, (5) \textbf{W}hy, describing the hazardous behavior or the trigger conditions that cause such behavior, (6) \textbf{H}ow, detailing the attack strategy the attacker utilized to trick the EVCS into being in a hazardous event situation. 

Every time a cyber incident occurs, the perpetrator (e.g., scammer, spy, terrorist, vandal, corporate thief, and skilled criminal) uses some platform (e.g., user instruction, script program, software suite, information tap, and data exchange) to carry out an attack or malicious activity by leveraging a flaw (e.g., architecture, execution, and settings) of a target component (e.g., procedure, information, charging infrastructure, and connectivity), resulting in an unauthorized outcome (e.g., data breach, resource theft, overcharging, and power outage) to meet their goals (e.g., political gain, financial gain, and damage). Therefore, it is crucial to investigate to ascertain the causes and implement remedies to stop such incidents from happening again. Therefore, the investigation team can utilize this technique to uncover trends in severe EVCS casualties and determine the root causes of these cyber occurrences. Please note that more advanced attack tools, techniques, and approaches in the system can be introduced in the future  (Table~\ref{tab:investigation} can be expanded with additional components).

Therefore, the main goal of the proposed work is to show how investigators can use the digital forensic model based on 5Ws \& 1H to examine the events that occurred in an EVCS incident in chronological order. Additionally, it will make it easier for an investigation team to have direction to take the appropriate steps.

\section{ANOMALY DETECTION IN A EVCS CYBERSECURITY INCIDENT}
EVCSs rely on a precise and reliable information exchange amongst several modules to follow a series of waypoints along a predetermined route in an environment. However, there are significant non-linearities associated with this information exchange, which are impacted by several aberrations and erroneous transmissions. A mishap could result from any transmission issue. As a result, a strategy based on stochastic M-ary classification is proposed to solve the issue of anomaly identification during event analysis in real-time cyber-attack scenarios. The probability of the various component operations is calculated using the probabilistic model suggested in the study, which also identifies anomalous behavior. It also calculates the likelihood of a cyber event based on probabilistic descriptions of different layer components in space and time.

\begin{figure}[t]
\centering
\includegraphics[width= 0.45\textwidth, height = 3.0 in]{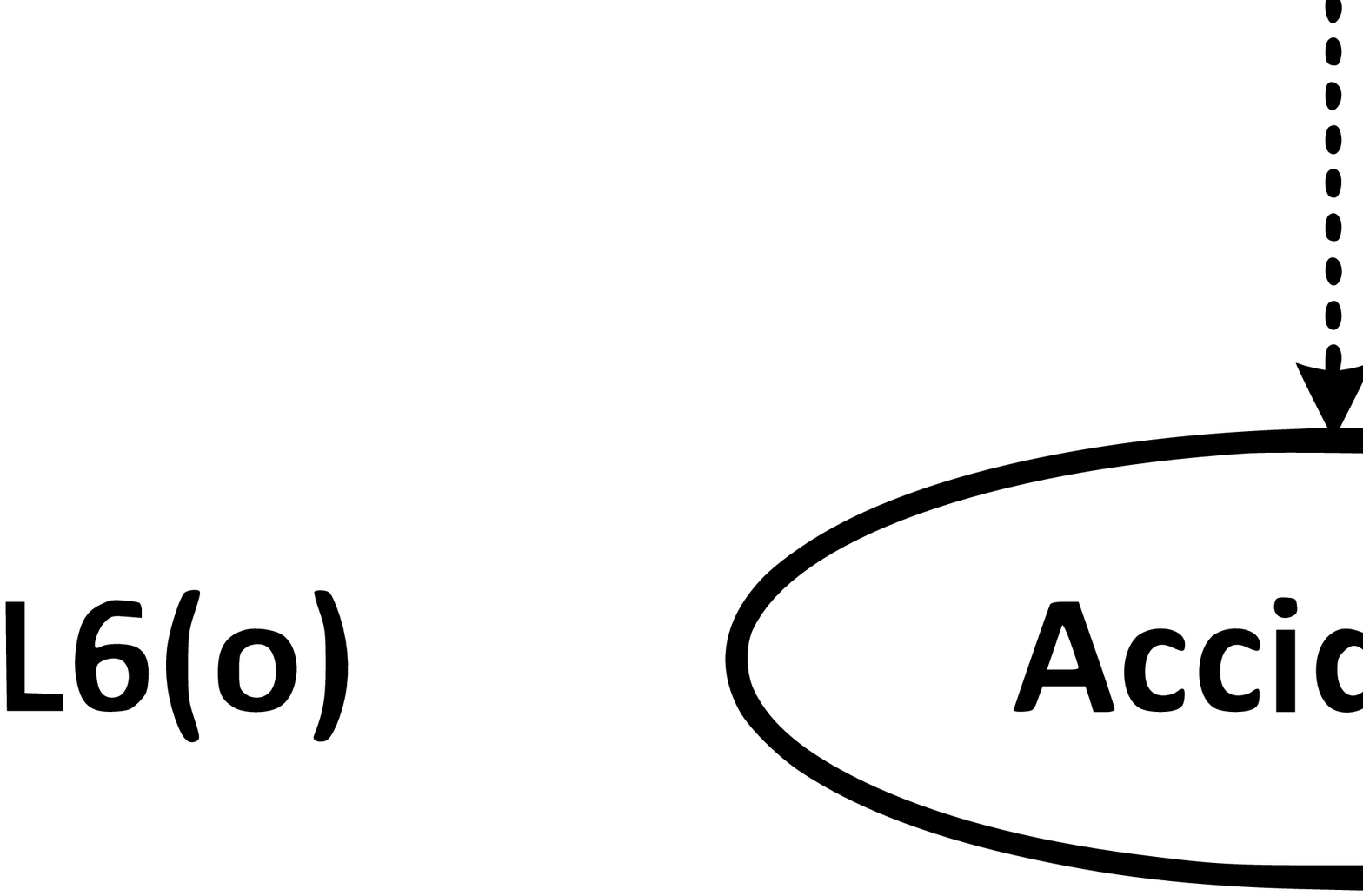}
\caption{Probability-based anomaly detection in a EVCS cyber incident.}
\label{fig:EVCS_algorithm}
\end{figure}

The networked structure model, as illustrated in Fig.~\ref{fig:EVCS_algorithm}, consists of six prime layers, which are explained as,\\
Every CPS is distinguished by a wired or wireless communication channel, which is further divided into various communication protocols supported by the EVCS environment, including OCPP, IEC 61850, ISO 15118, and others. Additionally, because the charging station environment is made up of three main core entities, these particular elements make up layer 3, their additional components comprise layer 4, and then their operations make up layer 5. Please note that each layer of this model can be adjusted to accommodate additional elements in the digital environment.

Mathematically, this model can be formulated to differentiate normal and abnormal behaviors such that,

Probability of occurrence of any abnormal event (E) due to incorrect transmissions of $A_{ec}$ is,

\begin{equation}
\text{P}(E)=\sum_{c=1}^{k}P(E \cap A_{ec}) = \sum_{c=1}^{k}P(E|A_{ec}) P(A_{ec}).
\end{equation}
In other words,
\begin{equation}
\begin{split}
    P(E|A_{ec}) =  P(E|A_{11})P(A_{11})+P(E|A_{12})P(A_{12})+ \\
    ...+P(E|A_{ec})P(A_{ec}),    
\end{split}
\end{equation}
where,
\begin{equation}
P(E|A_{11}) = \sum_{o=2}^{k}P(B_{co}|A_{11}).
\end{equation}
Summarize the above equation as,
\begin{equation}
P(E|A_{ec}) = \sum_{o=1}^{k} \sum_{c=1}^{k}P(B_{co}|A_{ec}), c\neq o.
\end{equation}
By Bayes theorem,
\begin{equation}
    P(B_{c2}|A_{11}) = \frac{P(A_{11}|B_{c2})P(B_{c2})}{P(A_{11})}.
\end{equation}
In general, 
\begin{equation}
    P(B_{co}|A_{ec}) = \frac{P(A_{ec}|B_{co})P(B_{co})}{P(A_{ec})}, c\neq o.
\end{equation}

The definitions to the symbols used in the above equations are explained as,
\\
P($A_{ec}$) = Probability that the transmitted signal from layer 3 is $A_{ec}$
\\
P($B_{co}$) = Probability that the received signal by layer 4 is $B_{co}$
\\
P($B_{co}$|$A_{ec}$) = Probability that the received signal is $B_{co}$, when the transmitted signal is $A_{c}$
\\
P(E|$A_{ec}$) = Probability of an abnormal behavior when $A_{ec}$ signal is transmitted layer 3\\
It is to be noted that all transmitted signals or messages ($A_{ec}$) are mutually exhaustive such that,
\begin{equation}
    \sum_{c=1}^{k}\sum_{e=1}^{k}P(A_{ec})=1.
\end{equation}
For the normal operation, c = o, and for error or abnormal behavior (E), c $\neq$ o.

\section{CASE STUDY}
\subsection{Attack Scenario 1}
\subsubsection{EVCS Cyber Incident Investigation}
The investigative team (or operator) arrives at the scene of the EVCS incident and gathers information from a variety of sources, including cloud servers and other sources, before establishing a correlation between the data. After that, reverse engineering is performed based on convincing evidence, and the analysis that follows may be given as indicated in Table~\ref{tab:Scenario1}.

\begin{table}[h!]
\caption{5Ws \& 1H-based investigation for attack scenario 1.}
\label{tab:Scenario1}
\begin{tabular}{|l|l|l|}
\hline
\multicolumn{1}{|c|}{\textbf{Attribute}} &
  \multicolumn{1}{c|}{\textbf{Definition}} &
  \multicolumn{1}{c|}{\textbf{Attack Scenario 1}} \\ \hline
\multirow{2}{*}{Who} &
  Attacker ($\alpha$) &
  $\alpha$: Hacker \\ \cline{2-3} 
 &
  Victim ($\nu$) &
  $\nu$: EVCS \\ \hline
What &
  \begin{tabular}[c]{@{}l@{}}Attack \\ target ($\tau$)\end{tabular} &
  $\tau$1: BMS, $\tau$2: CBs  \\ \hline
\multirow{2}{*}{When} &
  Date ($\delta$) &
  \begin{tabular}[c]{@{}l@{}}$\delta$1: 05, $\delta$2: 16, $\delta$3: 22, \\ Format: (05-16-22)\end{tabular} \\ \cline{2-3} 
 &
  Time ($\iota$) &
  \begin{tabular}[c]{@{}l@{}}$\iota$1: EST, $\iota$2: 02, $\iota$3: 20, $\iota$4: 40, \\ $\iota$5: 47, Format: (02:20:40:47)\end{tabular} \\ \hline
Where &
  \begin{tabular}[c]{@{}l@{}}Attack \\ path ($\rho$)\end{tabular} &
  \begin{tabular}[c]{@{}l@{}}$\rho$: OTA update \end{tabular} \\ \hline
Why &
  \begin{tabular}[c]{@{}l@{}}Hazardous \\ behavior ($\beta$)\end{tabular} &
  \begin{tabular}[c]{@{}l@{}}$\beta$1: False reporting, $\beta$2: flawed status \end{tabular} \\ \hline
How &
  \begin{tabular}[c]{@{}l@{}}Attack \\ method ($\omega$)\end{tabular} &
  $\omega$: Tampering \\ \hline
\end{tabular}
\end{table}
\subsubsection {Anomaly Detection}
\begin{figure}[t]
\centering
\includegraphics[width= 0.45\textwidth, height = 2.6 in]{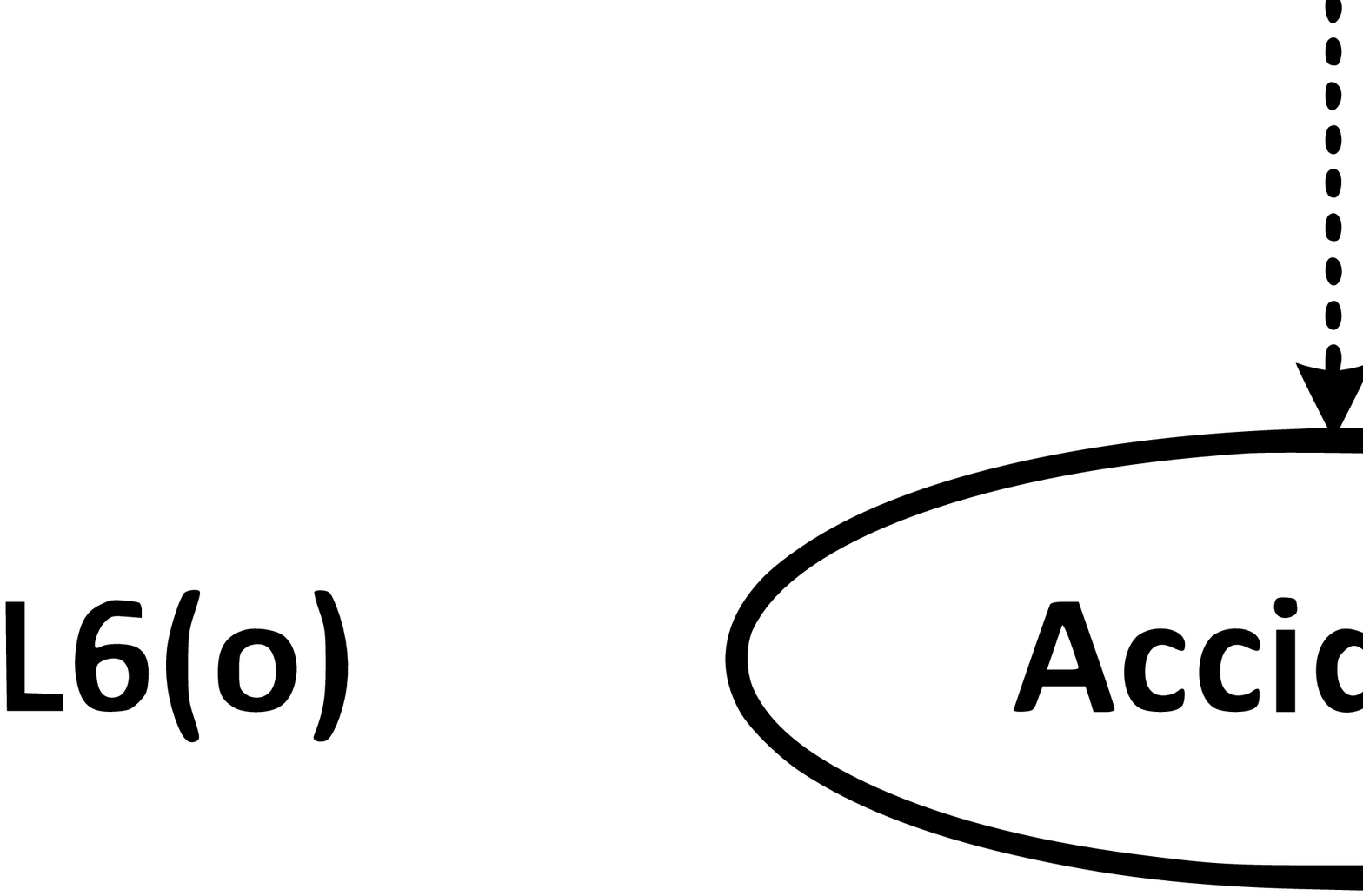}
\caption{Probability-based abnormal behavior analysis of attack scenario.}
\label{fig:EVCS_case}
\end{figure}

According to the proposed model, anomaly detection during the event analysis for the given scenario can be done as illustrated in Fig.~\ref{fig:EVCS_case}. In this situation, it is assumed that the BMS operates in charging mode, which means that it receives power from the utility and that the point of common coupling (PCC) breaker connecting the utility from the station is closed. However, overcurrent protection trips the PCC breaker the moment there is an external failure on the distribution feeder line while EVCS is in grid-connected operation. Therefore, to ensure a fully functional EVCS, the EVCS operator would open the CB and change the BMS's operating mode to discharging mode simultaneously. So, the anticipated outcome is that the BMS should report its changed mode of operation from charging to discharging, and the CB should report its changed status from closed to open to the EVCS. But because of abnormalities or cyber-attacks, the BMS displays an inaccurate charging schedule, and the CB displays an incorrect status, which leads to anomalous station behaviors.

\section{CONCLUSION}
An EVCS's modern environment merges the grid infrastructure, EVs, and other facilities. These include cloud servers, charging infrastructure, and communication protocols. The EVCS ecosystem is therefore driven by numerous interactions among complex cyber and physical systems. However, new technologies have also introduced new system vulnerabilities that may result in security breaches and could be an attractive target for adversaries, such as unauthorized remote access to EVCSs through misconfigured security devices (e.g., firewalls).
Additionally, the EVCS event is more complex due to the juxtaposition of several cyber aspects, making the investigation very difficult. Therefore, the efforts of 5Ws \& 1H-based digital investigative framework is proposed that can pinpoint the cybersecurity incidents in charge of the functional failure. Additionally, a robust probability-based anomaly detection algorithm is also designed to identify the function failure during a cyber event analysis.
The methods have been validated by testing with realistic intrusion scenario. For instnace, a case study is provided to support the proposed models, demonstrating their validity and dependability. Since the system lacks the real-time dataset to identify the attackers, the investigation model does have certain restrictions. Researchers may create distinct cyber-attack scenarios in the station environment and investigate them for future work utilizing these frameworks. The core causes of unusual station entity actions can be identified using these frameworks to detect a cyber event and investigate an incident.

\bibliographystyle{IEEEtran}

IEEEexample.bib
V1.12 (2007/01/11)
Copyright (c) 2002-2007 by Michael Shell
See: http://www.michaelshell.org/
for current contact information.

This is an example BibTeX database for the official IEEEtran.bst
BibTeX style file.

Some entries call strings that are defined in the IEEEabrv.bib file.
Therefore, IEEEabrv.bib should be loaded prior to this file.
Usage:

\bibliographystyle{./IEEEtran}
\bibliography{./IEEEabrv,./IEEEexample}


Support sites:
http://www.michaelshell.org/tex/ieeetran/
http://www.ctan.org/tex-archive/macros/latex/contrib/IEEEtran/
and/or
http://www.ieee.org/

*************************************************************************
Legal Notice:
This code is offered as-is without any warranty either expressed or
implied; without even the implied warranty of MERCHANTABILITY or
FITNESS FOR A PARTICULAR PURPOSE!
User assumes all risk.
In no event shall IEEE or any contributor to this code be liable for
any damages or losses, including, but not limited to, incidental,
consequential, or any other damages, resulting from the use or misuse
of any information contained here.

All comments are the opinions of their respective authors and are not
necessarily endorsed by the IEEE.

This work is distributed under the LaTeX Project Public License (LPPL)
( http://www.latex-project.org/ ) version 1.3, and may be freely used,
distributed and modified. A copy of the LPPL, version 1.3, is included
in the base LaTeX documentation of all distributions of LaTeX released
2003/12/01 or later.
Retain all contribution notices and credits.
** Modified files should be clearly indicated as such, including  **
** renaming them and changing author support contact information. **

File list of work: IEEEabrv.bib, IEEEfull.bib, IEEEexample.bib,
                   IEEEtran.bst, IEEEtranS.bst, IEEEtranSA.bst,
                   IEEEtranN.bst, IEEEtranSN.bst, IEEEtran_bst_HOWTO.pdf
*************************************************************************


Note that, because the example references were taken from actual IEEE
publications, these examples do not always contain the full amount
of information that may be desirable (for use with other BibTeX styles).
In particular, full names (not abbreviated with initials) should be
entered whenever possible as some (non-IEEE) bibliography styles use
full names. IEEEtran.bst will automatically abbreviate when it encounters
full names.

@book{hacker,
  author        = "Hein Smith and Hilary Morrison",
  title         = "Ethical Hacking: A Comprehensive Beginner’s Guide to Learn and Master Ethical Hacking",
  edition       = "Comprehensive",
  publisher     = "CreateSpace Independent Publishing Platform",
  year          = "{Jun. 21, 2018}"
}

@ARTICLE{6681962,
  author={D. M. E. {Ingram} and P. {Schaub} and R. R. {Taylor} and D. A. {Campbell}},
  journal=IEEE_J_PWRD, 
  title={System-Level Tests of Transformer Differential Protection Using an {IEC}61850 Process Bus}, 
  year={Jun. 2014},
  volume={29},
  number={3},
    pages={1382-1389},}

@ARTICLE{7752958,
  author={G. {Liang} and S. R. {Weller} and J. {Zhao} and F. {Luo} and Z. Y. {Dong}},
  journal={IEEE Transactions on Power Systems}, 
  title={The 2015 Ukraine Blackout: Implications for False Data Injection Attacks}, 
  year={2017},
  volume={32},
  number={4},
  pages={3317-3318},
  doi={10.1109/TPWRS.2016.2631891}}

@ARTICLE{4652578,
  author={C. {Ten} and C. {Liu} and G. {Manimaran}},
  journal={IEEE Transactions on Power Systems}, 
  title={Vulnerability Assessment of Cybersecurity for SCADA Systems}, 
  year={2008},
  volume={23},
  number={4},
  pages={1836-1846},
  doi={10.1109/TPWRS.2008.2002298}}

@ARTICLE{8930619,
  author={H. {Tu} and H. {Feng} and S. {Srdic} and S. {Lukic}},
  journal={IEEE Transactions on Transportation Electrification}, 
  title={Extreme Fast Charging of Electric Vehicles: A Technology Overview}, 
  year={2019},
  volume={5},
  number={4},
  pages={861-878},
  doi={10.1109/TTE.2019.2958709}}
  
  @INPROCEEDINGS{9265953,  
  author={J. {Straub}},  
  booktitle={2020 IEEE International Conference on Smart Cloud (SmartCloud)},  title={Modeling Attack, Defense and Threat Trees and the Cyber Kill Chain, ATT amp;CK and STRIDE Frameworks as Blackboard Architecture Networks},   year={2020},  
  volume={},  
  number={},  
  pages={148-153},  
  doi={10.1109/SmartCloud49737.2020.00035}}
  
  @ARTICLE{5477189,  
  author={Ten, Chee-Wooi and Manimaran, Govindarasu and Liu, Chen-Ching},  journal={IEEE Transactions on Systems, Man, and Cybernetics - Part A: Systems and Humans},   
  title={Cybersecurity for Critical Infrastructures: Attack and Defense Modeling},   
  year={2010},  
  volume={40},  
  number={4},  
  pages={853-865},  
  doi={10.1109/TSMCA.2010.2048028}}
  
  @ARTICLE{9123897,  
  author={Lee, Yousik and Woo, Samuel and Song, Yunkeun and Lee, Jungho and Lee, Dong Hoon},  
  journal={IEEE Access},   
  title={Practical Vulnerability-Information-Sharing Architecture for Automotive Security-Risk Analysis},   
  year={2020},  
  volume={8},  
  number={},  
  pages={120009-120018},  
  doi={10.1109/ACCESS.2020.3004661}}
  
  
  @ARTICLE{8788508,  
  author={Halvorsen, James and Waite, Jesse and Hahn, Adam},  
  journal={IEEE Access},   
  title={Evaluating the Observability of Network Security Monitoring Strategies With TOMATO},   
  year={2019},  
  volume={7},  
  number={},  
  pages={108304-108315}, 
  doi={10.1109/ACCESS.2019.2933415}}
  
  @ARTICLE{5477189,  
  author={C. {Ten}  and G. {Manimaran} and C. {Liu}},  
  journal={IEEE Transactions on Systems, Man, and Cybernetics - Part A: Systems and Humans},   
  title={Cybersecurity for Critical Infrastructures: Attack and Defense Modeling},   
  year={2010},  
  volume={40},  
  number={4},  
  pages={853-865},  
  doi={10.1109/TSMCA.2010.2048028}}
  
  @INPROCEEDINGS{9330666,  
  author={S. {Saadat} and S. {Maingot} and S. {Bahizad}},  
  booktitle={2020 5th IEEE Workshop on the Electronic Grid (eGRID)},   title={Electric Vehicle Charging Station Security Enhancement Measures},   year={2020},  
  volume={},  
  number={},  
  pages={1-8},  
  doi={10.1109/eGRID48559.2020.9330666}}

@INPROCEEDINGS{5638899,
  author={D. {Aggeler} and F. {Canales} and H. {Zelaya-De La Parra} and A. {Coccia} and N. {Butcher} and O. {Apeldoorn}},
  booktitle={2010 IEEE PES Innovative Smart Grid Technologies Conference Europe (ISGT Europe)}, 
  title={Ultra-fast DC-charge infrastructures for EV-mobility and future smart grids}, 
  year={2010},
  volume={},
  number={},
  pages={1-8},
  doi={10.1109/ISGTEUROPE.2010.5638899}}

@ARTICLE{8307044,
  author={S. {Habib} and M. M. {Khan} and F. {Abbas} and L. {Sang} and M. U. {Shahid} and H. {Tang}},
  journal={IEEE Access}, 
  title={A Comprehensive Study of Implemented International Standards, Technical Challenges, Impacts and Prospects for Electric Vehicles}, 
  year={2018},
  volume={6},
  number={},
  pages={13866-13890},
  doi={10.1109/ACCESS.2018.2812303}}

  
@ARTICLE{6678803,
  author={L. {Zhu} and D. {Shi} and P. {Wang}},
  journal=IEEE_J_PWRD, 
  title={{IEC}61850-Based Information Model and Configuration Description of Communication Network in Substation Automation}, 
  year={Feb. 2014},
  volume={29},
  number={1},
  pages={97-107},}  

@inproceedings{IEC61850-9,
  author        = "{IEC61850-9-2:2020 CSV}",
  title         = "Consolidated version, Communication networks and systems for power utility automation",
  booktitle     = "Part 9-2: Specific communication service mapping (SCSM) - Sampled values over ISO/IEC8802-3",
  address       = "",
  month         = "",
  year          = "2020",
  pages         = ""
  }
  
  @INPROCEEDINGS{9369743,  author={Girdhar, Mansi and Hong, Junho and Karnati, Ramya and Lee, Soonwoo and Choi, Sungsoo},  booktitle={2021 International Conference on Electronics, Information, and Communication (ICEIC)},   title={Cybersecurity of Process Bus Network in Digital Substations},   year={2021},  volume={},  number={},  pages={1-6},  doi={10.1109/ICEIC51217.2021.9369743}}
  
  @INPROCEEDINGS{9490069,  author={Sanghvi, Anuj and Markel, Tony},  booktitle={2021 IEEE Transportation Electrification Conference & Expo (ITEC)},   title={Cybersecurity for Electric Vehicle Fast-Charging Infrastructure},   year={2021},  volume={},  number={},  pages={573-576},  doi={10.1109/ITEC51675.2021.9490069}}
  
  @ARTICLE{9583592,  author={Girdhar, Mansi and Hong, Junho and Lee, Hyojong and Song, Tai-Jin},  journal={IEEE Transactions on Smart Grid},   title={Hidden Markov Models-Based Anomaly Correlations for the Cyber-Physical Security of EV Charging Stations},   year={2022},  volume={13},  number={5},  pages={3903-3914},  doi={10.1109/TSG.2021.3122106}}
  
  @INPROCEEDINGS{7785783,  author={Valdes, Alfonso and Macwan, Richard and Backes, Matthew},  booktitle={2016 IEEE 17th International Conference on Information Reuse and Integration (IRI)},   title={Anomaly Detection in Electrical Substation Circuits via Unsupervised Machine Learning},   year={2016},  volume={},  number={},  pages={500-505},  doi={10.1109/IRI.2016.74}}
  
  @INPROCEEDINGS{8493213,  author={Mültin, Marc},  booktitle={2018 International Conference of Electrical and Electronic Technologies for Automotive},   title={ISO 15118 as the Enabler of Vehicle-to-Grid Applications},   year={2018},  volume={},  number={},  pages={1-6},  doi={10.23919/EETA.2018.8493213}}
  
  @ARTICLE{9272723,  author={Acharya, Samrat and Dvorkin, Yury and Pandžić, Hrvoje and Karri, Ramesh},  journal={IEEE Access},   title={Cybersecurity of Smart Electric Vehicle Charging: A Power Grid Perspective},   year={2020},  volume={8},  number={},  pages={214434-214453},  doi={10.1109/ACCESS.2020.3041074}}
  
  @ARTICLE{9336258,  author={Wang, Lu and Qin, Zian and Slangen, Tim and Bauer, Pavol and van Wijk, Thijs},  journal={IEEE Open Journal of Power Electronics},   title={Grid Impact of Electric Vehicle Fast Charging Stations: Trends, Standards, Issues and Mitigation Measures - An Overview},   year={2021},  volume={2},  number={},  pages={56-74},  doi={10.1109/OJPEL.2021.3054601}}
  
  @ARTICLE{9786788,  author={Ahmad, Adnan and Qin, Zian and Wijekoon, Thiwanka and Bauer, Pavol},  journal={IEEE Open Journal of the Industrial Electronics Society},   title={An Overview on Medium Voltage Grid Integration of Ultra-Fast Charging Stations: Current Status and Future Trends},   year={2022},  volume={3},  number={},  pages={420-447},  doi={10.1109/OJIES.2022.3179743}}
 
@inproceedings{GDOI,
  author        = "{B. Weiss, M. Seewald, and H. Falk}",
  title         = "{GDOI} Protocol Support for {IEC} 62351 Security Services",
  booktitle     = "IETF",
  address       = "",
  month         = "June",
  year          = "2015",
  pages         = ""
  }  
  
  @INPROCEEDINGS{9215386,  author={Poornesh, Kavuri and Nivya, Kuzhivila Pannickottu and Sireesha, K.},  booktitle={2020 International Conference on Smart Electronics and Communication (ICOSEC)},   title={A Comparative study on Electric Vehicle and Internal Combustion Engine Vehicles},   year={2020},  volume={},  number={},  pages={1179-1183},  doi={10.1109/ICOSEC49089.2020.9215386}}
  
@inproceedings{IEC61850-9-2LE,
  author        = "{UCA International Users Group}",
  title         = "{IEC} 61850-9-2 {LE}: Implementation Guideline for Digital Interface to Instrument Transformer Using {IEC} 61850-9-2",
  booktitle     = "",
  address       = "",
  month         = "July",
  year          = "2004",
  pages         = ""
  }  

@inproceedings{IEC61850-1,
  author        = "{IEC61850-1}",
  title         = "Part 1: Introduction and overview",
  booktitle     = "Communication networks and systems for power utility automation",
  address       = "",
  month         = "Mar.",
  year          = "2013",
  pages         = ""
  }
  
@inproceedings{IEC62351-6ed1,
  author        = "{IEC62351-6:2007}",
  title         = "Power systems management and associated information exchange - Data and communication security",
  booktitle     = "Part 6: Security for IEC 61850",
  address       = "",
  month         = "",
  year          = "2007",
  pages         = ""
  }
  
@inproceedings{IEC62351-6ed2,
  author        = "{IEC62351-6:2020 PRV}",
  title         = "Power systems management and associated information exchange - Data and communication security",
  booktitle     = "Part 6: Security for IEC61850",
  address       = "",
  month         = "",
  year          = "2020",
  pages         = ""
  }  
  
@ARTICLE{PAC,
  author={F. {Hohlbaum} and M. {Braendle} and F. {Alvarez}},
  journal={PAC World Conf.}, 
  title={Cyber security practical considerations for implementing {IEC}62351}, 
  year={2010},
  volume={},
  number={},
  pages={1068-1078},}

@inproceedings{rtac,
  author        = "",
  title         = "{SEL}-3555 {R}eal-{T}ime {A}utomation {C}ontroller ({RTAC})",
  booktitle     = "",
  address       = "Available at: https://selinc.com/products/3555/",
  month         = "",
  year          = "",
  pages         = " "
  }
  
  @inproceedings{libiec,
  author        = "",
  title         = "Open source libraries for {IEC}61850",
  booktitle     = "",
  address       = "Available at: libiec61850.com",
  month         = "",
  year          = "",
  pages         = " "
  }

@ARTICLE{7536155,
  author={Q. {Huang} and S. {Jing} and J. {Li} and D. {Cai} and J. {Wu} and W. {Zhen}},
  journal=IEEE_J_PWRD, 
  title={Smart Substation: State of the Art and Future Development}, 
  year={April 2017},
  volume={32},
  number={2},
  pages={1098-1105},}


@inproceedings{PAC2010,
  author        = "F. Hohlbaum and M. Braendle and F. Alvarez",
  title         = "Cyber security practical considerations for implementing {IEC}62351",
  booktitle     = "PAC World Conf.",
  address       = "Dublin, Ireland",
  month         = "Jun.",
  year          = "2010",
  pages         = " "
  }
@ARTICLE{7372486,
  author={Y. {Zhang} and L. {Wang} and Y. {Xiang} and C. {Ten}},
  journal=IEEE_J_PWRS, 
  title={Inclusion of SCADA Cyber Vulnerability in Power System Reliability Assessment Considering Optimal Resources Allocation}, 
  year={Nov. 2016},
  volume={31},
  number={6},
  pages={4379-4394},}

@ARTICLE{9044460,
  author={J. {Wang} and G. {Constante} and C. {Moya} and J. {Hong}},
  journal={IET Cyber-Physical Systems: Theory   Applications}, 
  title={Semantic analysis framework for protecting the power grid against monitoring-control attacks}, 
  year={March 2020},
  volume={5},
  number={1},
  pages={119-126},}

@ARTICLE{8740995,
  author={S. M. S. {Hussain} and S. M. {Farooq} and T. S. {Ustun}},
  journal={IEEE Access}, 
  title={Analysis and Implementation of Message Authentication Code (MAC) Algorithms for GOOSE Message Security}, 
  year={June 2019},
  volume={7},
  number={},
  pages={80980-80984},}

@ARTICLE{7828156,
  author={T. T. {Tesfay} and J. {Le Boudec}},
  journal=IEEE_J_SG, 
  title={Experimental Comparison of Multicast Authentication for Wide Area Monitoring Systems}, 
  year={Sept. 2018},
  volume={9},
  number={5},
  pages={4394-4404},}

@ARTICLE{8918111,
  author={S. M. S. {Hussain} and T. S. {Ustun} and A. {Kalam}},
  journal=IEEE_J_IINF, 
  title={A Review of {IEC}62351 Security Mechanisms for {IEC}61850 Message Exchanges}, 
  year={Sept. 2020},
  volume={16},
  number={9},
  pages={5643-5654},}

@INPROCEEDINGS{8440438,
  author={D. {Ishchenko} and R. {Nuqui}},
  booktitle={2018 IEEE/PES Transmission and Distribution Conference and Exposition (T D)}, 
  title={Secure Communication of Intelligent Electronic Devices in Digital Substations}, 
  year={2018},
  volume={},
  number={},
  pages={1-5},}
  
@INPROCEEDINGS{7778772,
  author={M. {Strobel} and N. {Wiedermann} and C. {Eckert}},
  booktitle={2016 IEEE International Conference on Smart Grid Communications (SmartGridComm)}, 
  title={Novel weaknesses in {IEC}62351 protected Smart Grid control systems}, 
  year={2016},
  volume={},
  number={},
  pages={266-270},}  

@ARTICLE{9306781,
  author={R. {Zhu} and C. -C. {Liu} and J. {Hong} and J. {Wang}},
  journal={IEEE Access}, 
  title={Intrusion Detection Against MMS-Based Measurement Attacks at Digital Substations}, 
  year={2021},
  volume={9},
  number={},
  pages={1240-1249},
  doi={10.1109/ACCESS.2020.3047341}}

@ARTICLE{7553505,
  author={Y. {Yang} and H. {Xu} and L. {Gao} and Y. {Yuan} and K. {McLaughlin} and S. {Sezer}},
  journal={IEEE Transactions on Power Delivery}, 
  title={Multidimensional Intrusion Detection System for IEC 61850-Based SCADA Networks}, 
  year={2017},
  volume={32},
  number={2},
  pages={1068-1078},
  doi={10.1109/TPWRD.2016.2603339}}

@ARTICLE{8006250,
  author={J. {Hong} and C. {Liu}},
  journal={IEEE Transactions on Smart Grid}, 
  title={Intelligent Electronic Devices With Collaborative Intrusion Detection Systems}, 
  year={2019},
  volume={10},
  number={1},
  pages={271-281},
  doi={10.1109/TSG.2017.2737826}}

@ARTICLE{6786500,
  author={J. {Hong} and C. {Liu} and M. {Govindarasu}},
  journal={IEEE Transactions on Smart Grid}, 
  title={Integrated Anomaly Detection for Cyber Security of the Substations}, 
  year={2014},
  volume={5},
  number={4},
  pages={1643-1653},
  doi={10.1109/TSG.2013.2294473}}

@ARTICLE{7553505,
  author={Y. {Yang} and H. {Xu} and L. {Gao} and Y. {Yuan} and K. {McLaughlin} and S. {Sezer}},
  journal=IEEE_J_PWRD, 
  title={Multidimensional Intrusion Detection System for {IEC}61850-Based SCADA Networks}, 
  year={April 2017},
  volume={32},
  number={2},
  pages={1068-1078},}

@ARTICLE{9361308,
  author={J. {Hong} and R. {Karnati} and C. -W. {Ten} and S. {Lee} and S. {Choi}},
  journal={IEEE Transactions on Power Delivery}, 
  title={Implementation of Secure Sampled Value (SeSV) Messages in Substation Automation System}, 
  year={2021},
  volume={},
  number={},
  pages={1-1},
  doi={10.1109/TPWRD.2021.3061205}}

@ARTICLE{4753673,
  author={A. N. {Bessani} and P. {Sousa} and M. {Correia} and N. F. {Neves} and P. {Veríssimo}},
  journal={IEEE Security   Privacy}, 
  title={The Crutial Way of Critical Infrastructure Protection}, 
  year={2008},
  volume={6},
  number={6},
  pages={44-51},
  doi={10.1109/MSP.2008.158}}
  
@ARTICLE{8556464,
  author={J. {Hong} and R. F. {Nuqui} and A. {Kondabathini} and D. {Ishchenko} and A. {Martin}},
  journal=IEEE_J_IINF, 
  title={Cyber Attack Resilient Distance Protection and Circuit Breaker Control for Digital Substations}, 
  year={July 2019},
  volume={15},
  number={7},
  pages={4332-4341},}  
  
@ARTICLE{6530634,
  author={I. {Lim} and T. S. {Sidhu}},
  journal=IEEE_J_PWRD, 
  title={Design of a Backup IED for {IEC}61850-Based Substation}, 
  year={Oct. 2013},
  volume={28},
  number={4},
  pages={2048-2055},}
  
  
@ARTICLE{8006250,
  author={J. {Hong} and C. {Liu}},
  journal=IEEE_J_SG, 
  title={Intelligent Electronic Devices With Collaborative Intrusion Detection Systems}, 
  year={Jan. 2019},
  volume={10},
  number={1},
  pages={271-281},}  
  
 @ARTICLE{8234640,
  author={R. {Wójtowicz} and R. {Kowalik} and D. D. {Rasolomampionona}},
  journal=IEEE_J_PWRD, 
  title={Next Generation of Power System Protection Automation—Virtualization of Protection Systems}, 
  year={Aug. 2018},
  volume={33},
  number={4},
  pages={2002-2010},}
  
@ARTICLE{7898408,
  author={S. {Kariyawasam} and A. D. {Rajapakse} and N. {Perera}},
  journal=IEEE_J_PWRD, 
  title={Investigation of Using {IEC}61850-Sampled Values for Implementing a Transient-Based Protection Scheme for Series-Compensated Transmission Lines}, 
  year={Feb. 2018},
  volume={33},
  number={1},
  pages={93-101},}  

@INPROCEEDINGS{916941,
  author={D. C. {Elizondo} and J. {de La Ree} and A. G. {Phadke} and S. {Horowitz}},
  booktitle={2001 IEEE Power Engineering Society Winter Meeting. Conference Proceedings (Cat. No.01CH37194)}, 
  title={Hidden failures in protection systems and their impact on wide-area disturbances}, 
  year={2001},
  volume={2},
  number={},
  pages={710-714 vol.2},}




@ARTICLE{8097030,
    title   = "Toward Threat of Implementation Attacks on Substation Security: Case Study on Fault Detection and Isolation",
    author  = "Anupam Chattopadhyay and Abhisek Ukil and Dirmanto Jap and Shivam Bhasin",
    journal = IEEE_J_IINF,
    month   = "June",
    year    = "2018",
    volume  = "14",
    number  = "6",
    pages   = "2442-2451",
}

@ARTICLE{6630114,
  author={N. {Machiels} and N. {Leemput} and F. {Geth} and J. {Van Roy} and J. {Büscher} and J. {Driesen}},
  journal={IEEE Transactions on Smart Grid}, 
  title={Design Criteria for Electric Vehicle Fast Charge Infrastructure Based on Flemish Mobility Behavior}, 
  year={2014},
  volume={5},
  number={1},
  pages={320-327},
  doi={10.1109/TSG.2013.2278723}}

@ARTICLE{7469388,
  author={S. {Rivera} and B. {Wu}},
  journal={IEEE Transactions on Power Electronics}, 
  title={Electric Vehicle Charging Station With an Energy Storage Stage for Split-DC Bus Voltage Balancing}, 
  year={2017},
  volume={32},
  number={3},
  pages={2376-2386},
  doi={10.1109/TPEL.2016.2568039}}

@INPROCEEDINGS{5638899,
  author={D. {Aggeler} and F. {Canales} and H. {Zelaya-De La Parra} and A. {Coccia} and N. {Butcher} and O. {Apeldoorn}},
  booktitle={2010 IEEE PES Innovative Smart Grid Technologies Conference Europe (ISGT Europe)}, 
  title={Ultra-fast DC-charge infrastructures for EV-mobility and future smart grids}, 
  year={2010},
  volume={},
  number={},
  pages={1-8},
  doi={10.1109/ISGTEUROPE.2010.5638899}}

@ARTICLE{6472082,
  author={S. {Bai} and S. M. {Lukic}},
  journal={IEEE Transactions on Power Electronics}, 
  title={Unified Active Filter and Energy Storage System for an MW Electric Vehicle Charging Station}, 
  year={2013},
  volume={28},
  number={12},
  pages={5793-5803},
  doi={10.1109/TPEL.2013.2245146}}

@ARTICLE{7807218,
  author={D. T. {Hoang} and P. {Wang} and D. {Niyato} and E. {Hossain}},
  journal={IEEE Access}, 
  title={Charging and Discharging of Plug-In Electric Vehicles (PEVs) in Vehicle-to-Grid (V2G) Systems: A Cyber Insurance-Based Model}, 
  year={2017},
  volume={5},
  number={},
  pages={732-754},
  doi={10.1109/ACCESS.2017.2649042}}
  
  @Article{8870734,
    author={B. {Xu} and Z. {Zhong} and G.{He}},
    title={A Minimum Defense Cost Calculation Method for Attack Defense Trees},
    journal = {Security and Communication Networks},
    volume= {2020},
    year= 2020, 
    pages={1-12}, 
    doi={10.1155/2020/8870734}}
    
@ARTICLE{8813023,  
author={Avatefipour, Omid and Al-Sumaiti, Ameena Saad and El-Sherbeeny, Ahmed M. and Awwad, Emad Mahrous and Elmeligy, Mohammed A. and Mohamed, Mohamed A. and Malik, Hafiz},  
journal={IEEE Access},   
title={An Intelligent Secured Framework for Cyberattack Detection in Electric Vehicles’ CAN Bus Using Machine Learning},   
year={2019},  
volume={7},  
number={},  
pages={127580-127592},  
doi={10.1109/ACCESS.2019.2937576}}   

@article{100013,
author = {Dimitrios Kosmanos and Apostolos Pappas and Leandros Maglaras and Sotiris Moschoyiannis and Francisco J. Aparicio-Navarro and Antonios Argyriou and Helge Janicke},
title = {A novel Intrusion Detection System against spoofing attacks in connected Electric Vehicles},
journal = {Array},
volume = {5},
pages = {1-11},
year = {2020},
doi = {https://doi.org/10.1016/j.array.2019.100013}}

@ARTICLE{8788508,  
author={Halvorsen, James and Waite, Jesse and Hahn, Adam},  
journal={IEEE Access},   
title={Evaluating the Observability of Network Security Monitoring Strategies With TOMATO},   
year={2019},  
volume={7},  
number={},  
pages={108304-108315},  
doi={10.1109/ACCESS.2019.2933415}}

@ARTICLE{9123897,  
author={Lee, Yousik and Woo, Samuel and Song, Yunkeun and Lee, Jungho and Lee, Dong Hoon},  
journal={IEEE Access},   
title={Practical Vulnerability-Information-Sharing Architecture for Automotive Security-Risk Analysis},   
year={2020},  
volume={8},  
number={},  
pages={120009-120018},  
doi={10.1109/ACCESS.2020.3004661}}

@INPROCEEDINGS{9243152,  
author={Basnet, Manoj and Hasan Ali, Mohd.},  
booktitle={2nd International Conference on Smart Power   Internet Energy Systems (SPIES)},   
title={Deep Learning-based Intrusion Detection System for Electric Vehicle Charging Station},   
year={2020},  
volume={},  
number={},  
pages={408-413},  
doi={10.1109/SPIES48661.2020.9243152}}
    
@ARTICLE{8667410,  
author={Li, Qianmu and Meng, Shunmei and Wang, Shuo and Zhang, Jing and Hou, Jun},  
journal={IEEE Access},   
title={CAD: Command-Level Anomaly Detection for Vehicle-Road Collaborative Charging Network},   
year={2019},  
volume={7},  
number={},  
pages={34910-34924},  
doi={10.1109/ACCESS.2019.2904047}}


@INPROCEEDINGS{4275642,  
author={Ten, Chee-Wooi and Liu, Chen-Ching and Govindarasu, Manimaran},  
booktitle={2007 IEEE Power Engineering Society General Meeting},   
title={Vulnerability Assessment of Cybersecurity for SCADA Systems Using Attack Trees},   
year={2007},  
volume={},  
number={},  
pages={1-8},  
doi={10.1109/PES.2007.385876}}


@INPROCEEDINGS{8377181,
  author={Y. {Fraiji} and L. {Ben Azzouz} and W. {Trojet} and L. A. {Saidane}},
  booktitle={2018 IEEE Wireless Communications and Networking Conference (WCNC)}, 
  title={Cyber security issues of Internet of electric vehicles}, 
  year={2018},
  volume={},
  number={},
  pages={1-6},
  doi={10.1109/WCNC.2018.8377181}}

@ARTICLE{8030362,
  author={A. {Bindra}},
  journal={IEEE Power Electronics Magazine}, 
  title={Securing the Power Grid: Protecting Smart Grids and Connected Power Systems from Cyberattacks}, 
  year={2017},
  volume={4},
  number={3},
  pages={20-27},
  doi={10.1109/MPEL.2017.2719201}}

@ARTICLE{9272723,
  author={S. {Acharya} and Y. {Dvorkin} and H. {Pandžić} and R. {Karri}},
  journal={IEEE Access}, 
  title={Cybersecurity of Smart Electric Vehicle Charging: A Power Grid Perspective}, 
  year={2020},
  volume={8},
  number={},
  pages={214434-214453},
  doi={10.1109/ACCESS.2020.3041074}}

@ARTICLE{8106743,
    title   = "Anomaly Detection Based on Zone Partition for Security Protection of Industrial Cyber-Physical Systems",
    author  = "Jun Yang and Chunjie Zhou and Shuanghua Yang and Haizhou Xu and Bowen Hu",
    journal = IEEE_J_IINF,
    month   = "May",
    year    = "2018",
    volume  = "65",
    number  = "5",
    pages   = "4257-4267",
}

@ARTICLE{8556464,
  author        = "Junho Hong and Reynaldo Nuqui and Anil Kondabathini and Dmitry Ishchenko and Aaron Martin",
  title         = "Cyber Attack Resilient Distance Protection and Circuit Breaker Control for Digital Substations",
  journal       = IEEE_J_IINF,
  doi           = {10.1109/TII.2018.2884728},
  note          = "(to appear)"
}

@inproceedings{CRAM1,
  author        = "Z. Yang and C.-W. Ten",
  title         = "Cyber-Induced Risk Modeling for Microprocessor-Based Relays in Substations",
  booktitle     = "Proc. 2018 IEEE Conf. Innov. Smart Grid Technol.--Asia (ISGT--Asia)",
  address       = "Singapore",
  month         = "May",
  year          = "2018",
  pages         = "856-861"  
  }

@MISC{new1,
  author        = "{US-CERT Alert (TA18-074A)}",
  title         = "Russian Government Cyber Activity Targeting Energy and Other Critical Infrastructure Sectors",
  month         = "Mar. 15",
  url           = "https://www.us-cert.gov/ncas/alerts/TA18-074A",
  year          = "2018",
}

@MISC{new2,
  author        = "{ICS-CERT}",
  title         = "MAR-17-352-01 HatMan-Safety System Targeted Malware (Update A)",
  month         = "Apr. 10",
  url           = "https://ics-cert.us-cert.gov/sites/default/files/documents/MAR-17-352-01\%20HatMan\%20-\%20Safety\%20System\%20Targeted\%20Malware\%20\%28Update\%20A\%29_S508C.PDF",
  year          = "2018",
}

@book{PMAnderson,
  author        = "P. M. Anderson",
  title         = "Power System Protection",
  edition       = "1",
  publisher     = "The Institute of Electrical and Electronics Engineers, Inc.",
  address       = "New York, USA"
  year          = "1998"
}

@ARTICLE{Cascade3,
    title   = "Risk Assessment of Cascading Outages: Methodologies and Challenges",
    author  = "M. Vaiman and K. Bell and Y. Chen and B. Chowdhury and I. Dobson and P. D. H. Hines and M. Papic and S. Miller and P. Zhang",
    journal = IEEE_J_PWRS,
    month   = "May",
    year    = "2012",
    volume  = "27",
    number  = "2",
    pages   = "631-641",
}

@ARTICLE{Cascade2,
    title   = "Expose hidden failures to prevent cascading outages",
    author  = "A. G. Phadke and J. S. Thorp",
    journal = "IEEE Comput. Appl. Power ",
    month   = "Jul.",
    year    = "1996",
    volume  = "9",
    number  = "3",
    pages   = "20-23",
}

@ARTICLE{Cascade1,
    title   = "A Loading-Dependent Model of Probabilistic Cascading Failure ",
    author  = "Ian Dobson  and Benjamin A. Carreras  and David E. Newman ",
    journal = "Probab. Eng. Inf. Sci.",
    month   = "Jan.",
    year    = "2005",
    volume  = "19",
    number  = "1",
    pages   = "15-32",
}

@book{DigRelay3,
  author        = "A. T. Johns  and  S. K. Salman",
  title         = "Digital Protection for Power Systems",
  publisher     = "Institution of Engineering and Technology",
  series = {Energy Engineering},
  year          = " 1995"
}

@ARTICLE{DigRelay2,
    title   = "A DIGITAL MULTIFUNCTION PROTECTIVE RELAY ",
    author  = "Murty V. V. S. Yalla",
    journal = IEEE_J_PWRD,
    month   = "Jan.",
    year    = "1992",
    volume  = "7",
    number  = "1",
    pages   = "193-201",
}

@ARTICLE{DigRelay1,
    title   = "Digital protective relaying algorithm sensitivity study and evaluation",
    author  = "M. Kezunovic and S. Kreso and J. T. Cain and B. Perunicic",
    journal = IEEE_J_PWRD,
    month   = "Jul.",
    year    = "1988",
    volume  = "3",
    number  = "3",
    pages   = "912-922",
}

@electronic{ERCOT,
  author        = "{Electric Reliability Council of Texas, Inc.}",
  title         = "{ERCOT} Nodal Operating Guides, Section 2: system Operations and Control Requirements",
  address       = " ",
  url           = "http://www.ercot.com/content/wcm/current_guides/53525/02-060118.doc",
  year          = " Jun. 1, 2018"
  }

@electronic{IEEE14,
  author        = "{Power Systems Test Case Archive}",
  title         = "14 Bus Power Flow Test Case",
  address       = "University of Washington",
  url           = "https://www2.ee.washington.edu/research/pstca/pf14/pg_tca14bus.htm",
  year          = " Aug., 1993"
  }

@book{IEC8,
  author        = "{IEC61850-8-1}",
  title         = "{Communication networks and systems for power utility automation--Part 8--1: Specific communication service mapping (SCSM)--Mappings to MMS
(ISO 9506-1 and ISO 9506-2) and to ISO/IEC8802-3}",
  edition       = "2.0",
}

@book{IEC9,
  author        = "{IEC61850-9-2}",
  title         = "{Communication Networks And Systems For Power Utility Automation - Part 9-2: Specific Communication Service Mapping (SCSM) - Sampled Values Over ISO/IEC8802-3}",
  edition       = "2.0",
  publisher     = "{IEC}",
  month         = "",
  year          = " 2011"
}

@book{IEC7,
  author        = "{IEC61850-7-4 Edition 2.0}",
  title         = "{Communication Networks and Systems for Power Utility Automation Part 7-4: Basic Communication Structure--Compatible logical node classes and data object classes}"
}

@electronic{FERC1,
  author        = "{Federal Energy Regulatory Commission (FERC)}",
  title         = "Cyber Security Incident Reporting Reliability Standards",
  address       = "888 First Street, NE Washington, DC",
  url           = "https://www.ferc.gov/whats-new/comm-meet/2018/071918/E-1.pdf",
  year          = " Jul. 19, 2018"
  }

@book{CPAT,
  author        = "{Power System Stability Study Group}",
  title         = "Integrated Analysis Software for Bulk Power System Stability",
  publisher     = "ET90002",
  address       = "CRIEPI Report",
  year          = "Jul. 1991"
  }
  
@Book{relayguide,
 author    = "{Meter, Relay and Instrucment Division}",
 title     = "Protective Relays Application Guide",
 publisher = "The English Electric Company Limited",
 year      =  "1968",
 edition   = "First"
}

@ARTICLE{Cybersecurity1,
    title   = "Cyber-Physical System Security of a Power Grid: State-of-the-Art",
    author  = "Chih-Che Sun and Chen-Ching Liu and Jing Xie",
    journal = " Electronics",
    month   = "Jul.",
    year    = "2016",
    volume  = "5",
    number  = "3",
    pages   = "1-18",
}
@book{18535,
author        = "{National Research Council}",
editor        = "David W. Cooke",
title         = "The Resilience of the Electric Power Delivery System in Response to Terrorism and Natural Disasters: Summary of a Workshop",
address       = "Washington, DC",
year          = "2013",
publisher     = "The National Academies Press"
}

@book{24836,
author        = "{National Academies of Sciences, Engineering, and Medicine}",
title         = "Enhancing the Resilience of the Nation's Electricity System",
address       = "Washington, DC",
year          = "2017",
publisher     = "The National Academies Press"
}

@inproceedings{digitalrelay3,
  author        = "S. Ward and et al.",
  title         = "Cyber Security Issues for Protective Relays; C1 Working Group Members of Power System Relaying Committee",
  booktitle     = "Proc. 2007 IEEE Power Eng. Soc. General Meeting",
  address       = " Tampa, FL, USA",
  month         = "Jun.",
  year          = "2007",
  pages         = "1-8"
  }


@techreport{digitalrelay2,
    AUTHOR = {Kenneth C. Behrendt},
    TITLE = {Relay-to-Relay Digital Logic Communication
for Line Protection, Monitoring, and Control},
    INSTITUTION = {Schweitzer Engineering Laboratories, Inc.},
    ADDRESS = {Pullman, Washington {USA}},
    YEAR = {1998},
    MONTH = {Nov.},
    URL = {https://cdn.selinc.com/assets/Literature/Publications/Technical\%20Papers/6029_RelayToRelay_Web.pdf}
}

@ARTICLE{digitalrelay1,
    title   = "Microprocessor-based protective relays",
    author  = "R. D. Kirby and R. A. Schwartz",
    journal = "IEEE Ind. Appl. Mag.",
    month   = "Sep.",
    year    = "2009",
    volume  = "15",
    number  = "5",
    pages   = "43-50",
}

@ARTICLE{IEEEstd,
    title   = "{IEEE} Guide for Power System Protection Testing",
    author  = "{IEEE Power \& Energy Society} ",
    journal = "IEEE Std C37.233-2009",
    month   = "Dec.",
    year    = "2009",
    volume  = " ",
    number  = " ",
    pages   = "1-112",
}

@ARTICLE{relaytest,
    title   = "Real-time software testing for microprocessor-based protective relays",
    author  = "N. I. Santoso and J. Y. Avins",
    journal = IEEE_J_PWRD,
    month   = "Jul.",
    year    = "1994",
    volume  = "9",
    number  = "3",
    pages   = "1359-1367",
}


@inproceedings{modernprotection,
  author        = "L. Ayers and M. Lanier and L. Wright",
  title         = "Protecting distribution substation assets -- Modern protection schemes with microprocessor-based relays",
  booktitle     = "Proc. 2013 66th Annu. Conf. Protective Relay Engineers",
  address       = " College Station, TX, USA",
  month         = "Apr.",
  year          = "2013",
  pages         = "1-7"
  }


@ARTICLE{wirelessrelay,
    title   = "A Practical Layered Multiplexed-Coded Relaying Scheme for Wireless Multicast",
    author  = "K. Mazher and U. B. Farooq and J. N. Chattha and M. Uppal",
    journal = "IEEE Trans. Veh. Technol." ,
    month   = "Jan.",
    year    = "2018",
    volume  = "64",
    number  = "1",
    pages   = "554-566",
}


@ARTICLE{travelingwave,
    title   = "Protective Relays with Traveling Wave Technology Revolutionize Fault Locating",
    author  = "E. O. Schweitzer and A. Guzman and M. V. Mynam and V. Skendzic and B. Kasztenny and S. Marx",
    journal = "IEEE Power Energy Mag." ,
    month   = "Mar.",
    year    = "2016",
    volume  = "14",
    number  = "2",
    pages   = "114-120",
}

@article{905746,
  author        = "T. Krause and R. Ernst and B. Klaer and I. Hacker and M. Henze",
  title         = "Cybersecurity in Power Grids: Challenges and Opportunities",
  journal       = "arXiv",
  address       ="",
  volume        = "",
  number  ="",
  year          = "2021",
  pages         = ""
}

@INPROCEEDINGS{9916758,  author={Hong, Junho and Girdhar, Mansi and Ten, Chee-Wooi and Lee, Soonwoo and Choi, Sungsoo},  booktitle={2022 IEEE Power & Energy Society General Meeting (PESGM)},   title={Cybersecurity of Sampled Value Messages in Substation Automation System},   year={2022},  volume={},  number={},  pages={1-1},  doi={10.1109/PESGM48719.2022.9916758}}

@INPROCEEDINGS{9698889,  author={Hafeez, Azeem and Mohan, Janani and Girdhar, Mansi and Awad, Selim S.},  booktitle={2021 17th International Computer Engineering Conference (ICENCO)},   title={Machine Learning based ECU Detection for Automotive Security},   year={2021},  volume={},  number={},  pages={73-81},  doi={10.1109/ICENCO49852.2021.9698889}}

@article{KORONIOTIS202091,
title = {A new network forensic framework based on deep learning for Internet of Things networks: A particle deep framework},
journal = {Future Generation Computer Systems},
volume = {110},
pages = {91-106},
year = {2020},
issn = {0167-739X},
doi = {https://doi.org/10.1016/j.future.2020.03.042},
url = {https://www.sciencedirect.com/science/article/pii/S0167739X19325105},
author = {Nickolaos Koroniotis and Nour Moustafa and Elena Sitnikova},
}

@ARTICLE{9849671,  author={Girdhar, Mansi and You, Yongsik and Song, Tai-Jin and Ghosh, Subhadip and Hong, Junho},  journal={IEEE Access},   title={Post-Accident Cyberattack Event Analysis for Connected and Automated Vehicles},   year={2022},  volume={10},  number={},  pages={83176-83194},  doi={10.1109/ACCESS.2022.3196346}}

@INPROCEEDINGS{9243536,  author={Mohd Abdullah, Haris Iskandar and Mustaffa, Muhammad Zulhusni and Rahim, Fiza Abdul and Ibrahim, Zul-Azri and Yusoff, Yunus and Yussof, Salman and Bakar, Asmidar Abu and Ismail, Roslan and Ramli, Ramona},  booktitle={2020 8th International Conference on Information Technology and Multimedia (ICIMU)},   title={Smart Grid Digital Forensics Investigation Framework},   year={2020},  volume={},  number={},  pages={200-205},  doi={10.1109/ICIMU49871.2020.9243536}}

@article{905fg578,
  author        = "Y. Yusoff, R. Ismail and Z. Hassan",
  title         = "COMMON PHASES OF COMPUTER FORENSICS INVESTIGATION MODELS",
  journal       = "International Journal of Computer Science & Information Technology (IJCSIT)",
  address       ="",
  volume        = "3",
  number        = "3",
  year          = "2011",
  pages         = ""
}

@ARTICLE{ZhaoandZhang2017a,
    title   = "Observing individual dynamic choices of activity chains from location-based crowdsourced data",
    author  = "Shuaidong Zhao and Kuilin Zhang",
    journal = "Transportation Res. Part C: Emerging Technol." ,
    month   = "Dec.",
    year    = "2017",
    volume  = "85",
    number  = "1",
    pages   = "1-22",
}

@ARTICLE{Hagerstrand1970,
    title   = "What About People in Regional Science? ",
    author  = "T. H{\"a}gerstrand",
    journal = "Papers of the Regional Sci.  Assoc. " ,
    month   = " ",
    year    = "1970",
    volume  = "24",
    number  = "1",
    pages   = "6-21",
}

@ARTICLE{Miller1991,
    title   = "Modelling accessibility using space-time prism concepts within geographical information systems",
    author  = "Harvey J. Miller",
    journal = "Int. J. of Geographical Inform.  Syst. " ,
    month   = " ",
    year    = "1991",
    volume  = "5",
    number  = "3",
    pages   = "287-301",
}

@ARTICLE{Katz1960,
    title   = "THE FUNCTIONAL APPROACH TO THE STUDY OF ATTITUDES",
    author  = "Daniel Katz",
    journal = "Public Opinion Quart. " ,
    month   = "Jan.",
    year    = "1960",
    volume  = "24",
    number  = "2",
    pages   = "163-204",
}

@inproceedings{Zhangetal2015,
  author        = "Y. Zhang and S. Shen and J. L. Mathieu",
  title         = "Data-driven optimization approaches for optimal power flow with uncertain reserves from load control",
  booktitle     = "Proc. 2015 Amer. Control Conf. (ACC)",
  address       = "Chicago, IL, USA",
  month         = "Jul.",
  year          = "2015",
  pages         = "3013-3018"
  }

@ARTICLE{KangandLee2017,
    title   = "Data-Driven Optimization of Incentive-based Demand Response System with Uncertain Responses of Customers",
    author  = "Jimyung Kang and Jee-Hyong Lee",
    journal = "Energies" ,
    month   = "Oct.",
    year    = "2017",
    volume  = "10",
    number  = "10",
    pages   = "1996-1073",
}

@ARTICLE{Xuetal2018,
    title   = "Data-Driven Pricing Strategy for Demand-Side Resource Aggregators",
    author  = "Z. Xu and T. Deng and Z. Hu and Y. Song and J. Wang",
    journal = IEEE_J_SG ,
    month   = "Jan.",
    year    = "2018",
    volume  = "9",
    number  = "1",
    pages   = "57-66",
}

@inproceedings{ZhaoandZhang2017b,
  author        = "Shuaidong Zhao and Kuilin Zhang",
  title         = "A Data-Driven Optimization Model to Observe Individual Dynamic Choices of Activity-Travel-Path using Connected Vehicles as Mobile Sensors",
  booktitle     = "Proc. The 96th Annu.  Meeting of Transportation Res.  Board",
  address       = "Washington D.C., USA",
  month         = "Jan.",
  year          = "2017",
  pages         = " "
  }
@inproceedings{ZhaoandZhang2017c,
  author        = "Shuaidong Zhao and Kuilin Zhang",
  title         = "Observing Space-Time Queueing Dynamics at a Signalized Intersection using Vehicles as Mobile Sensors",
  booktitle     = "Proc. The 96th Annu.  Meeting of Transportation Res.  Board",
  address       = "Washington D.C., USA",
  month         = "Jan.",
  year          = "2017",
  pages         = " "
  }
@inproceedings{ZhaoandZhang2018,
  author        = "Shuaidong Zhao and Kuilin Zhang",
  title         = "A data-driven Model Predictive Control framework for robust Cooperative Adaptive Cruise Control using mobile sensing data",
  booktitle     = "Proc. The 97th Annu.  Meeting of Transportation Res.  Board",
  address       = "Washington D.C., USA",
  month         = "Jan.",
  year          = "2018",
  pages         = " "
  }
  
@book{Scarf1958,
  author        = "H. SCARF",
  title         = "In Studies in The Mathematical Theory of Inventory and Production",
  series        = "A min-max solution of an inventory problem",
  publisher     = "Stanford University Press",
  edition       = "{K. Arrow, S. Karlin and H. Scarf}",
  address       = "California",
  year          = "1958",
  pages         = "201-209"
}

@ARTICLE{BertsimasPopescu2005,
    title   = "Optimal Inequalities in Probability Theory: A Convex Optimization Approach",
    author  = "Dimitris Bertsimas and Ioana Popescug",
    journal = "SIAM J. Optim." ,
    month   = " ",
    year    = "2005",
    volume  = "15",
    number  = "3",
    pages   = "780-804",
}

@ARTICLE{DelageEYeY2010,
    title   = "Distributionally Robust Optimization Under Moment Uncertainty with Application to Data-Driven Problems",
    author  = "Erick Delage and Yinyu Ye",
    journal = "Oper. Res." ,
    month   = "Jan. ",
    year    = "2010",
    volume  = "58",
    number  = "3",
    pages   = "595-612",
}

@ARTICLE{BenTaletal2013,
    title   = "Robust Solutions of Optimization Problems Affected by Uncertain Probabilities",
    author  = "Aharon Ben-Tal and Dick den Hertog and Anja De Waegenaere and Bertrand Melenberg and Gijs Rennen",
    journal = "Manage. Sci. " ,
    month   = "Nov. ",
    year    = "2012",
    volume  = "59",
    number  = "2",
    pages   = "341 - 357",
}
@ARTICLE{Wangetal2015,
    title   = "Likelihood robust optimization for data-driven problems",
    author  = "Zizhuo Wang and Peter W. Glynn and Yinyu Ye",
    journal = "Computational Manage. Sci. " ,
    month   = "Apr. ",
    year    = "2016",
    volume  = "13",
    number  = "2",
    pages   = "241-261",
}
@ARTICLE{EsfahaniPMKuhnD2017,
    title   = "Data-Driven Distributionally Robust Optimization using the Wasserstein Metric: Performance Guarantees and Tractable Reformulations",
    author  = "Peyman Mohajerin Esfahani and Daniel Kuhn",
    journal = "Math. Programming " ,
    month   = "Jul. ",
    year    = "2017",
    volume  = "1",
    number  = "1",
    pages   = "1-52",
}





@inproceedings{IED2,
  author        = "Junho Hong and Chen-Ching Liu and Manimaran Govindarasu",
  title         = "Detection of cyber intrusions using network-based multicast messages for substation automation",
  booktitle     = "Innovative Smart Grid Technologies (ISGT), 2014 IEEE PES",
  month         = "Feb.",
  year          = "2014",
  pages         = "1-5"
  }

@ARTICLE{IED1,
    title   = "Intelligent Electronic Devices with Collaborative Intrusion Detection Systems",
    author  = "J. Hong and C. C. Liu",
    journal = IEEE_J_SG,
    month   = " to be published",
}


@ARTICLE{STS1,
    title   = "Intersection Numbers of Kirkman Triple Systems",
    author  = "Yanxun Chang and Giovanni Lo Faro",
    journal = "J.  Combinatorial Theory, Series A" ,
    month   = "Jan.",
    year    = "1998",
    volume  = "86",
    number  = "2",
    pages   = "348-361",
}

@ARTICLE{STS2,
    title   = "EFrames for Kirkman triple systems",
    author  = "D. R. Stinson",
    journal = "Discrete Math." ,
    month   = "Jul.",
    year    = "1987",
    volume  = "65",
    number  = "3",
    pages   = "289-300",
}


@electronic{Laplacian1,
  author        = "Yousef Saad",
  title         = "Iterative Methods for Sparse Linear Systems",
  address       = "Society for Industrial and Applied Mathematics",
  url           = "http://www-users.cs.umn.edu/~saad/IterMethBook_2ndEd.pdf",
  year          = "2003"
  }

@Book{Designtheory,
 author    = "Jeffrey H. Dinitz and Douglas R. Stinson",
 title     = "Contemporary Design Theory: A collection of Surveys",
 publisher = "A Wiley-Interscience Publication ",
 year      =  1992,
 edition   = "First"
}

@ARTICLE{Extended,
    title   =  "Extended Enumeration of Hypothesized Substations Outages Incorporating Overload Implication",
    author  =  "Z. Yang and C.-W. Ten and A. Ginter",
    journal =  IEEE_J_SG,
    month   =  "Nov.",
    year    =  "2018",
    volume  =  "9",
    number  =  "6",
    pages   =  "6929-6938",
}

@ARTICLE{redundant,
    title   = "Substation reliability evaluation including switching actions with redundant components",
    author  = "J. J. Meeuwsen and W. L. Kling",
    journal = IEEE_J_PWRD ,
    month   = "Oct.",
    year    = "1997",
    volume  = "12",
    number  = "4",
    pages   = "1472-1479",
}

@Book{whitepaper,
 author    = "Manimaran Govindarasu and Adam Hahn and Peter Sauer",
 title     = "Cyber-Physical Systems Security for Smart Grid: Future Grid Initiative White Paper",
 publisher = "(Power Systems Engineering Research Center) PSERC",
 year      =  2012,
 edition   = "First"
}

@MISC{Shodan,
    author  =  "{Cable News Network (CNN)}",
    title   = "Shodan: The scariest search engine on the Internet",
    month   = "Apr. 8",
    year    = "2013",
    Url     = {http://money.cnn.com/2013/04/08/technology/security/shodan/index.html},
}

@MISC{Nmap,
    author  =  "{NMAP.ORG}",
    title   = "Chapter 15. Nmap Reference Guide",
    month   = "Mar. 25",
    year    = "2011",
    Url     = {https://nmap.org/book/man.html},
}

@MISC{Wireshark,
    author  =  "Richard Sharpe and Ed Warnicke",
    title   = "Wireshark User�s Guide",
    month   = "Nov. 9",
    year    = "2014",
    Url     = {https://www.wireshark.org/docs/wsug_html/},
}


@MISC{Wannacry1,
    author  =  "{Cable News Network (CNN)}",
    title   = "WannaCrypt ransomware attack should make us wanna cry",
    month   = "May 14",
    year    = "2017",
    Url     = {http://www.cnn.com/2017/05/14/opinions/wannacrypt-attack-should-make-us-wanna-cry-about-vulnerability-urbelis/index.html},
}

@MISC{Wannacry2,
    author  = "Henry Belot and Stephanie Borys",
    title   = "Ransomware attack still looms in Australia as Government warns WannaCry threat not over",
    month   = "May 16",
    year    = "2017",
    Url     = {http://www.abc.net.au/news/2017-05-15/ransomware-attack-to-hit-victims-in-australia-government-says/8526346},
}

@MISC{Ukraine1,
    author  = "{Electricity Information Sharing and Analysis Center (E-ISAC)}",
    title   = "Analysis of the Cyber Attack on the Ukrainian Power Grid",
    month   = "March 18",
    year    = "2016",
    Url     = {http://www.nerc.com/pa/CI/ESISAC/Documents/E-ISAC_SANS_Ukraine_DUC_18Mar2016.pdf},
}

@MISC{ICSCERT,
  author        = "{ICS-CERT Alert (IR-ALERT-H-16-056-01)}",
  title         = "Cyber-Attack Against Ukrainian Critical Infrastructure",
  month         = "Feb. 25",
  url           = "https://ics-cert.us-cert.gov/alerts/IR-ALERT-H-16-056-01",
  year          = "2016",
}

@ARTICLE{impact,
    title   =  "Impact Assessment of Hypothesized Cyberattacks on Interconnected Bulk Power Systems",
    author  =  "C.-W. Ten and K. Yamashita and Z. Yang and A. Vasilakos and A. Ginter",
    journal =  IEEE_J_SG,
    month   =  "Sep.",
    year    =  "2018",
    volume  =  "9",
    number  =  "5",
    pages   =  "4405-4425",
}

@MISC{nerc1,
    author  = "{{NERC} Board of Trustees}",
    title   = "Reliability Standards for the Bulk Electric Systems of North America",
    month   = "May",
    year    = "2017",
    Url     = {http://www.nerc.com/pa/Stand/Reliability\%20Standards\%20Complete\%20Set/RSCompleteSet.pdf},
}

@MISC{ANSSI1,
    author  = "{Agence nationale de la s\'ecurit\'e des syst\`emes d'information (ANSSI)}",
    title   = "Classification Method and Key Measures",
    month   = "Jan.",
    year    = "2014",
    Url     = {https://www.ssi.gouv.fr/uploads/2014/01/industrial_security_WG_Classification_Method.pdf},
}

@MISC{ANSSI2,
    author  = "{Agence nationale de la s\'ecurit\'e des syst\`emes d'information (ANSSI)}",
    title   = "Detailed Measures",
    month   = "Jan.",
    year    = "2014",
    Url     = {https://www.ssi.gouv.fr/uploads/2014/01/industrial_security_WG_detailed_measures.pdf},
}

@MISC{nist1,
    author  = " {National Institute of Standards and Technology (NIST)}",
    title   = "Improving Critical Infrastructure Cybersecurity Executive Order 13636: Preliminary Cybersecurity Framework ",
    month   = "Jun.",
    year    = "2013",
    Url     = {https://www.nist.gov/sites/default/files/documents/itl/preliminary-cybersecurity-framework.pdf},
}

@MISC{nist2,
    author  = " {National Institute of Standards and Technology (NIST)}",
    title   = "Cybersecurity Framework Workshop 2017 Summary",
    month   = "Jul.",
    year    = "2017",
    Url     = {https://www.nist.gov/sites/default/files/documents/2017/07/21/cybersecurity_framework_workshop_2017_summary_20170721_1.pdf},
}

@MISC{generator,
    author  =  "{Cable News Network (CNN)}",
    title   = "Sources: Staged cyber attack reveals vulnerability in power grid",
    month   = "Sep. 26",
    year    = "2007",
    Url     = {http://www.cnn.com/2007/US/09/26/power.at.risk/index.html?iref=topnews},
}

@MISC{Stuxnet,
    author  = "D. Veluz",
    title   = "Stuxnet Malware Targets {SCADA} Systems",
    month   = "Oct.",
    year    = "2010",
    Url     = {http://www.trendmicro.com/vinfo/us/threat-encyclopedia/web-attack/54/stuxnet-malware-targets-scada-systems},
}

@MISC{ukraine,
    author  = "Evan Perez",
    title   = "First on {CNN}: {U.S.} investigators find proof of cyberattack on {Ukraine} power grid",
    month   = "Feb. 3",
    year    = "2016",
    Url     = {http://www.cnn.com/2016/02/03/politics/cyberattack-ukraine-power-grid/},
}

@MISC{CIP,
    author  = "{Critical Infrastructure Protection Committee (CIPC)}",
    title   = "Cybersecurity -- BES Cyber System Categorization",
    month   = "Oct. 26",
    year    = "2012",
    Url     = {http://www.netsectech.com/wp-content/uploads/2013/05/Version-5-of-the-NERC-CIP-Cyber-Security-Standards.pdf},
}

@MISC{Ex.Ukraine,
  author        = "Kim Zetter",
  title         = "Everything We Know About Ukraine�s Power Plant Hack",
  month         = "Jan. 20",
  url           = "http://www.wired.com/2016/01/everything-we-know-about-ukraines-power-plant-hack/",
  year          = "2016",
}

@MISC{USAToday,
  author        = "Steve Reilly",
  title         = "{USA Today} Records: Energy Department struck by cyber attacks",
  month         = "Sep. 11",
  url           = "http://www.usatoday.com/story/news/2015/09/09/cyber-attacks-doe-energy/71929786/",
  year          = "2015",
}


@ARTICLE{ajjarapu,
    title   = "The continuation power flow: a tool for steady state voltage stability analysis",
    author  = "V. Ajjarapu and C. Christy",
    journal = IEEE_J_PWRS ,
    month   = "Feb.",
    year    = "1992",
    volume  = "7",
    number  = "1",
    pages   = "416-423",
}

@ARTICLE{8031986,
  author={P. {Holgado} and V. A. {Villagrá} and L. {Vázquez}},
  journal={IEEE Transactions on Dependable and Secure Computing}, 
  title={Real-Time Multistep Attack Prediction Based on Hidden Markov Models}, 
  year={2020},
  volume={17},
  number={1},
  pages={134-147},
  doi={10.1109/TDSC.2017.2751478}}
  
@ARTICLE{8031986,
  author={P. {Holgado} and V. A. {Villagrá} and L. {Vázquez}},
  journal={IEEE Transactions on Dependable and Secure Computing}, 
  title={Real-Time Multistep Attack Prediction Based on Hidden Markov Models}, 
  year={2020},
  volume={17},
  number={1},
  pages={134-147},
  doi={10.1109/TDSC.2017.2751478}}
  
@ARTICLE{DC1,
    title   = "On The Formulation of Power Distribution Factors for Linear Load Flow Methods",
    author  = "P. W. Sauer",
    journal = "IEEE Trans. Power App. Syst." ,
    month   = "Feb.",
    year    = "1981",
    volume  = "PAS-100",
    number  = "2",
    pages   = "764-700",
}

@ARTICLE{151351235,
    title   = "{DOE/DHS/DOT} volpe technical meeting on electric vehicle and charging station cybersecurity",
    author  = "K. Harnett, B. Harris, D. Chin, and G. Watson",
    journal = "U.S. Department of Transportation, Tech. Report. DOT-VNTSC-DOE-18-01" ,
    month   = "Mar.",
    year    = "2018",
    volume  = "",
    number  = "",
    pages   = "",
}

@ARTICLE{15657579,
    title   = "Electric Sector Failure Scenarios and Impact Analysis",
    author  = "",
    journal = "Electric Power Research Institute (EPRI), NESCOR Tech. Report" ,
    month   = "Dec.",
    year    = "2015",
    volume  = "",
    number  = "",
    pages   = "",
}


@ARTICLE{1231351351235,
    title   = "Symposium on federally funded research on cybersecurity of electric vehicle supply equipment (EVSE)",
    author  = "S. Lightman and T. Brewer",
    journal = "Nat. Inst. Standards Technol. NISTIR 8294" ,
    month   = "April",
    year    = "2020",
    volume  = "",
    number  = "",
    pages   = "",
}

@INPROCEEDINGS{9202653,
  author={P. {Danielis} and M. {Beckmann} and J. {Skodzik}},
  booktitle={2020 IEEE 44th Annual Computers, Software, and Applications Conference (COMPSAC)}, 
  title={An ISO-Compliant Test Procedure for Technical Risk Analyses of IoT Systems Based on STRIDE}, 
  year={2020},
  volume={},
  number={},
  pages={499-504},
  doi={10.1109/COMPSAC48688.2020.0-203}}

@INPROCEEDINGS{8260283,
  author={R. {Khan} and K. {McLaughlin} and D. {Laverty} and S. {Sezer}},
  booktitle={2017 IEEE PES Innovative Smart Grid Technologies Conference Europe (ISGT-Europe)}, 
  title={STRIDE-based threat modeling for cyber-physical systems}, 
  year={2017},
  volume={},
  number={},
  pages={1-6},
  doi={10.1109/ISGTEurope.2017.8260283}}

@INPROCEEDINGS{8260283,
  author={R. {Khan} and K. {McLaughlin} and D. {Laverty} and S. {Sezer}},
  booktitle={2017 IEEE PES Innovative Smart Grid Technologies Conference Europe (ISGT-Europe)}, 
  title={STRIDE-based threat modeling for cyber-physical systems}, 
  year={2017},
  volume={},
  number={},
  pages={1-6},
  doi={10.1109/ISGTEurope.2017.8260283}}

@ARTICLE{IEC61850,
    title   = "Automated analysis of power system events",
    author  = "V. Ajjarapu and C. Christy",
    journal = "IEEE Power Energy Mag." ,
    month   = "Sept.",
    year    = "2005",
    volume  = "3",
    number  = "5",
    pages   = "48-55",
}

@ARTICLE{9387290,  author={Amin, Md Ali Reza Al and Shetty, Sachin and Njilla, Laurent and Tosh, Deepak K. and Kamhoua, Charles},  journal={IEEE Access},   title={Hidden Markov Model and Cyber Deception for the Prevention of Adversarial Lateral Movement},   year={2021},  volume={9},  number={},  pages={49662-49682},  doi={10.1109/ACCESS.2021.3069105}}

@ARTICLE{12342351351,
    title   = "An integrated approach to safety and security based on systems theory",
    author  = "W. Young and N. G. Leveson",
    journal = "Communications of the ACM" ,
    month   = "Feb.",
    year    = "2014",
    volume  = "57",
    number  = "2",
    pages   = "",
}


@ARTICLE{Cyber1,
    title   = "Cyber Security and Power System Communication -- Essential Parts of a Smart Grid Infrastructure",
    author  = "G. N. Ericsson",
    journal = IEEE_J_PWRD,
    month   = "Jul.",
    year    = "2010",
    volume  = "25",
    number  = "3",
    pages   = "1501-1507",
}


@Article{island3,
  author    = "Qianchuan Zhao and Kai Sun and Da-Zhong Zheng and Jin Ma and Qiang Lu",
  title     = "A study of system splitting strategies for island operation of power system: a two-phase method based on OBDDs",
  journal   = IEEE_J_PWRS,
  month     = "Nov.",
  year      = "2003",
  volume    = "18",
  number    = "4",
  pages     = "1556-1565",
}


@ARTICLE{Dunstan,
    title   = "Digital Load Flow Studies",
    author  = "L. A. Dunstan",
    journal = IEEE_J_AIEE3,
    month   = "Jan.",
    year    = "1954",
    volume  = "73",
    number  = "1",
    pages   = "825-832",
}


@ARTICLE{FWu,
    title    = "Theoretical study of the convergence of the fast decoupled load flow",
    author   = "Felix F. Wu",
    journal  = IEEE_J_PWRAS,
    month    = "Jan.",
    year     = "1977",
    volume   = "96",
    number   = "1",
    pages    = "268-275",
}


@ARTICLE{Anomadetect,
    title    = "Anomaly Detection for Cybersecurity of the Substations",
    author   = "C.-W. Ten and J. Hong and C. C. Liu",
    journal  = IEEE_J_SG,
    month    = "Dec.",
    year     = "2011",
    volume   = "2",
    number   = "4",
    pages    = "865-873",
}

@ARTICLE{Subauto,
    title    = "Substation automation technologies and advantages",
    author   = "S. Bricker and T. Gonen and L. Rubin",
    journal  = IEEE_M_CAP,
    month    = "Jul.",
    year     = "2001",
    volume   = "14",
    number   = "3",
    pages    = "31-37",
}


@ARTICLE{1:realtimecontingency,
    title   = "Real-Time Contingency Analysis With Transmission Switching on Real Power System Data",
    author  = "M. Sahraei-Ardakani and X. Li and P. Balasubramanian and K. W. Hedman and M. Abdi-Khorsand",
    journal = IEEE_J_PWRS,
    month   = "May",
    year    = "2016",
    volume  = "31",
    number  = "3",
    pages   = "2501-2502",
}

@MISC{2:EOcommission,
    author    = "{Office of the Press Secretary, The White House}",
    title     = "Executive Order -- Commission on Enhancing National Cybersecurity",
    month     = "Feb. 9",
    year      = "2016",
    Url       = {https://www.whitehouse.gov/the-press-office/2016/02/09/executive-order-commission-enhancing-national-cybersecurity},
}

@MISC{3:EO13636,
    author    = "{Federal Register}",
    title     = "Executive Order 13636 -- Improving Critical Infrastructure Cybersecurity",
    month     = "Feb. 19",
    year      = "2013",
    Url       = "http://www.gsa.gov/portal/mediaId/176567/fileName/ATTCH\_1\_-\_CyberEO\-FedReg.action",
}


@Article{5:Cyberbase,
  author    = "C.-W. Ten and A. Ginter and R. Bulbul",
  title     = "Cyber-Based Contingency Analysis",
  journal   = IEEE_J_PWRS,
  month     = "Jul.",
  year      = "2016",
  volume    = "31",
  number    = "4",
  pages     = "3040--3050",
}

@MISC{6:review,
    author   = "LISA O. MONACO",
    title    = " Administration Efforts on Cybersecurity: The Year in Review and Looking Forward to 2016",
    month    = "Feb. 2",
    year     = "2016",
    Url      = {https://www.whitehouse.gov/blog/2016/02/02/administration-efforts-cybersecurity-year-review-and-looking-forward-2016},
}

@inproceedings{8:riskindex,
  author        = "R. Bulbul and Y. Gong and C.-W. Ten and A. Ginter and S. Mei",
  title         = "Impact quantification of hypothesized attack scenarios on bus differential relays",
  booktitle     = "Proc. IEEE Power Systems Computation Conference (PSCC)",
  address       = "Wroclaw, Poland",
  month         = "Aug.",
  year          = "2014",
  pages         = "1-7"
  }

@inproceedings{PLC,
  author        = "H. Wardak and S. Zhioua and A. Almulhem",
  title         = "{PLC} access control: a security analysis",
  booktitle     = "Proc. 2016 World Congress on Industrial Control Systems Security (WCICSS)",
  address       = "London, UK",
  month         = "Dec.",
  year          = "2016",
  pages         = "1-6"
  }


@ARTICLE{9:cybersecurity,
    title    = "Cybersecurity Myths on Power Control Systems: 21 Misconceptions and False Beliefs",
    author   = "L. Pi\`etre-Cambac\'ed\`es and M. Tritschler and G. N. Ericsson",
    journal  = IEEE_J_PWRD,
    month    = "Jan.",
    year     = "2011",
    volume   = "26",
    number   = "1",
    pages    = "161-172",
}


@inproceedings{11:dynamictopolgy,
  author        = "Mingyang Li and Qianchuan Zhao and P. B. Luh",
  title         = "{DC} power flow in systems with dynamic topology",
  booktitle     = "Proc. IEEE PES General Meeting",
  address       = "Pittsburgh, PA",
  month         = "Jul.",
  year          = "2008",
  pages         = "1-8"
  }

@ARTICLE{12:scpse,
    title    = "{SCPSE}: Security-Oriented Cyber-Physical State Estimation for Power Grid Critical Infrastructures",
    author   = "S. Zonouz and K. M. Rogers and R. Berthier and R. B. Bobba and W. H. Sanders and T. J. Overbye",
    journal  = IEEE_J_SG,
    month    = "Dec.",
    year     = "2012",
    volume   = "3",
    number   = "4",
    pages    = "1790-1799",
}

@Book{13:book,
 author    = "Federico Milano",
 title     = "Power System Modelling and Scripting",
 publisher = "Springer",
 year      =  2010,
 edition   = "First"
}



@ARTICLE{15:cascading2,
    title   = "Cascading Failure Analysis With {DC} Power Flow Model and Transient Stability Analysis",
    author  = "J. Yan and Y. Tang and H. He and Y. Sun",
    journal = IEEE_J_PWRS,
    month   = "Jan.",
    year    = "2015",
    volume  = "30",
    number  = "1",
    pages   = "285-297",
}

@ARTICLE{16:cascading3,
    title    = "A ``Random Chemistry'' Algorithm for Identifying Collections of Multiple Contingencies That Initiate Cascading Failure",
    author   = "M. J. Eppstein and P. D. H. Hines",
    journal  = IEEE_J_PWRS,
    month    = "Aug.",
    year     = "2012",
    volume   = "27",
    number   = "3",
    pages    = "1698-1705",
}

@ARTICLE{17:matpower,
    title   = "{MATPOWER}: Steady-State Operations, Planning, and Analysis Tools for Power Systems Research and Education",
    author  = "R. D. Zimmerman and C. E. Murillo-Sanchez and R. J. Thomas",
    journal = IEEE_J_PWRS,
    month   = "Feb.",
    year    = "2011",
    volume  = "26",
    number  = "1",
    pages   = "12-19",
}

@inproceedings{18:ramprate,
  author        = "D. Cormode and A. D. Cronin and W. Richardson and A. T. Lorenzo and A. E. Brooks and D. N. DellaGiustina",
  title         = "Comparing ramp rates from large and small {PV} systems, and selection of batteries for ramp rate control",
  booktitle     = "Proc. IEEE 39th Photovoltaic Specialists Conf. (PVSC)",
  address       = "Tampa, FL",
  month         = "June",
  year          = "2013",
  pages         = "1805-1810"
  }

  @inproceedings{20:standards,
  author        = "D. D. Roybal",
  title         = "Standards and ratings for the application of molded case, insulated case, and power circuit breakers",
  booktitle     = "Pulp and Paper Industry Technical Conference, 2000. Conference Record of 2000 Annual",
  address       = "Atlanta, GA, USA",
  month         = "June",
  year          = "2000",
  pages         = "195-204"
  }

  @inproceedings{21:computing,
  author        = "Z. Huang and Y. Chen and J. Nieplocha",
  title         = "Massive contingency analysis with high performance computing",
  booktitle     = "Power Energy Society General Meeting, 2009. PES '09. IEEE",
  address       = "Calgary, AB",
  month         = "July",
  year          = "2009",
  pages         = "1-8"
  }


@ARTICLE{23:overloads,
    title     = "Transmission-line overloads: real-time control",
    author    = "W. R. Lachs",
    journal   = "IEE Proceedings C - Generation, Transmission and Distribution",
    month     = "Sept.",
    year      = "1987",
    volume    = "134",
    number    = "5",
    pages     = "342-347",
}

@electronic{24:volstable,
  author        = "Varun Togit",
  title         = "Pattern Recognition of Power System Voltage Stability using Statistical and Algorithmic Methods",
  address       = "University of New Orleans",
  url           = "http://scholarworks.uno.edu/cgi/viewcontent.cgi.pdf",
  year          = "2012"
  }

@Book{25:book2,
 author    = "J.D. Glover and M. Sarma",
 title     = "Power system analysis and design�",
 publisher = "Thomson Learning",
 year      =  2008,
 edition   = "4th Edition"
}

@inproceedings{26:limits,
  author        = "Y. Zhao and A. Goldsmith and H. V. Poor",
  title         = "Fundamental limits of cyber-physical security in smart power grids",
  booktitle     = "Proc. IEEE 52nd Annual Conf. on Decision and Control (CDC)",
  address       = "Firenze",
  month         = "Dec.",
  year          = "2013",
  pages         = "200-205"
  }

@inproceedings{IEEECyberStand,
  author        = "{Power system relaying and substations committees}",
  title         = "{IEEE} standard cybersecurity requirements for substsation automation, protection, and control systems ",
  booktitle     = "IEEE PES Std C37.240-2014",
  address       = "IEEE-SA standards board",
  month         = "Dec.",
  year          = "2014",
  pages         = "1-38"
  }

@inproceedings{29:island2,
  author        = "R. Sun and Z. Wu and V. A. Centeno",
  title         = "Power system islanding detection \& identification using topology approach and decision tree",
  booktitle     = "Power and Energy Society General Meeting, 2011 IEEE",
  address       = "San Diego, CA",
  month         = "July",
  year          = "2011",
  pages         = "1-6"
  }

@ARTICLE{30:highvolt,
  author        = "Dominik Pieniazek",
  title         = "HV Substation Design:Applications and Considerations",
  url           = "http://www.hv-eng.com/IEEE\_CED\_SubDesign\_Full\_Size.pdf",
  year          = "2012",
}

@ARTICLE{31:island3,
    title    = "Two-Step Spectral Clustering Controlled Islanding Algorithm",
    author   = "L. Ding and F. M. Gonzalez-Longatt and P. Wall and V. Terzija",
    journal  = "IEEE Trans. Power Syst.",
    month    = "Feb.",
    year     = "2013",
    volume   = "28",
    number   = "1",
    pages    = "75-84",
}


@inproceedings{34:contingency1,
  author        = "M. A. C. Camargo and A. J. Rivera and R. R. Pe�a",
  title         = "Impact assessment of substation contingencies in power systems",
  booktitle     = "Transmission Distribution Conference and Exposition - Latin America (PES T D-LA), 2014 IEEE PES",
  address       = "Medellin",
  month         = "Sept.",
  year          = "2014",
  pages         = "1-6"
  }

@inproceedings{35:contingency2,
  author        = "S. H. Song and S. H. Lee and T. K. Oh and J. Lee",
  title         = "Risk-based contingency analysis for transmission and substation planning",
  booktitle     = "Proc. IEEE Trans. Dist. Conf. Exposition: Asia and Pacific",
  address       = "Seoul",
  month         = "Oct.",
  year          = "2009",
  pages         = "1-4"
  }


@electronic{cascading1,
  author        = "Mario Rios and Keith Bell and Daniel Kirschen and Ron Allan",
  title         = "Computation of the Value of Security",
  address       = "Mannchester Centre for Electrical Energy, UMIST",
  url           = "http://www2.ee.washington.edu/research/real/Library/Reports/Value_of_Security_Part_I.pdf",
  year          = "1999"
  }

@book{cascadingphd,
  author        = "Pooya Rezaei",
  title         = "Cascading Failure Risk Estimation and Mitigation in Power Systems",
  publisher     = "University of Vermont",
  address       = "Ph.D. dissertation",
  year          = "2015"
  }


@ARTICLE{Z3,
    title    = "Vulnerability assessment of cybersecurity for {SCADA} system",
    author="Chee-Wooi Ten and Chen-Ching Liu and Govindarasu Manimaran ",
    journal= IEEE_J_PWRS,
    month ="Nov.",
    year="2008",
    volume="23",
    number="4",
    pages="1836-1846",
}

@ARTICLE{Z4,
    title="Intrusion Evaluation of Communication Network Architectures for Power Substations",
    author="Rashiduzzaman Bulbul and Pingal Sapkota and Chee-Wooi Ten and Lingfeng Wang and Andrew Ginter",
    journal="IEEE Trans. Power Syst.",
    month ="June",
    year="2015"
    volume="30",
    number="30",
    pages="1372-1382",
}

@ARTICLE{Z5,
    title="Estimating a System's Mean Time-to-Compromise",
    author="Leversage, D.J. and James, E.",
    journal="IEEE Security Privacy",
    month ="Aug.",
    year="2008",
    volume="6",
    number="1",
    pages="52-60",
}



@book{Z6,
  author        = "Anany Levitin",
  title         = "Introduction to The Design and Analysis of Algorithm",
  publisher     = "Addison-Wesley",
  address       = "New Jersey",
  year          = "2011"
  }
@inproceedings{Z7,
  author        = "Li Fu and Wenjie Huang and Sheng Xiao and Yuan Li and Shifan Guo",
  title         = "Vulnerability Assessment for Power Grid Based on Small-world Topological Model",
  booktitle     = "Power and Energy Engineering Conference (APPEEC), 2010 Asia-Pacific",
  address       = "Rio de Janeiro, Brazil",
  month         = "Mar.",
  year          = "2010",
  pages         = "1-4"
  }
@ARTICLE{Z8,
    title="Identifying High risk N-k contingencies for on-line secrity assessment",
    author="Q. Chen and J. McCalley",
    journal="IEEE Trans. Power Syst.",
    month ="May",
    year="2005",
    volume="20",
    number="2",
    pages="823-834",
}
@electronic{Z9,
  author        = "Richard P. Stanley",
  title         = "Topics in Algebraic Combinatorics,Course notes for Mathematics 192",
  address       = "Harvard University",
  url           = "http://www-math.mit.edu/~rstan/algcomb/algcomb.pdf",
  year          = "2013"
  }

@ARTICLE{Z10,
    title="Modeling and Validation of Electrical Load Profiling in Residential Buildings in Singapore",
    author="C. Luo and U. Abhisek",
    journal="IEEE Trans. Power Syst.",
    month ="Sep.",
    year={2015},
    volume={30},
    number={5},
    pages={2800-2809},
}
@book{Z11,
  author        = "Andrea Sportiello",
  title         = "Combinatorial Methods in Statistical Field Theory",
   url           = "http://pcteserver.mi.infn.it/~caraccio/PhD/Sportiello.pdf",
  year          = "2003"
  }

  @book{Z12,
  author        = " J. Lewis Blackburn and Thomas J. Domin.",
  title         = "Protective relaying principles and applications ",
  publisher     = "Boca Raton, FL : CRC Press",
  url           = "http://pcteserver.mi.infn.it/~caraccio/PhD/Sportiello.pdf",
  year          = "2007"
  }


@ARTICLE{Z13,
    title="Contingency Analysis and Identification of Dynamic Voltage Control Areas",
    author="Paramasivam, M. and Dasgupta, S. and Ajjarapu, V. and Vaidya, U.",
    journal=IEEE_J_PWRS,
    month ="Nov.",
    year="2015",
    volume="30",
    number="6",
    pages="2974 - 2983",
    }

@ARTICLE{Z14,
    title="Cyber-Physical Modeling and Cyber-Contingency Assessment of Hierarchical Control Systems",
    author="Shujun Xin and Qinglai Guo and  Hongbin Sun and  Boming Zhang and  Jianhui Wang and Chen Chen",
    journal= IEEE_J_SG,
    month ="Sept.",
    year={2015},
    volume={6},
    number={5},
    pages={2375-2385},
}

@ARTICLE{Z15,
    title="A Framework for Modeling Cyber-Physical Switching Attacks in Smart Grid",
    author="Shan Liu and S. Mashayekh and D. Kundur and T. Zourntos and K. Butler-Purry.",
    journal= IEEE_J_ETC,
    month ="Dec.",
    year={2013},
    volume={1},
    number={2},
    pages={273-285},
}


@inproceedings{Z16,
  author        = "Ituzaro, F.A. and Douglin, R.H. and Butler-Purry, K.L.",
  title         = "Zonal overcurrent protection for smart radial distribution systems with distributed generation",
  booktitle     = "Innovative Smart Grid Technologies (ISGT), 2013 IEEE PES",
  month         = "Feb.",
  year          = "2013",
  pages         = "1-6"
  }


@ARTICLE{Z17,
    title="Synchronized Phasor Measurement Applications in Power Systems",
    author="De La Ree, J. and Centeno, V. and Thorp, J.S. and Phadke, A.G.",
    journal=IEEE_J_SG,
    month ="May",
    year={2010},
    volume={1},
    number={1},
    pages={20-27},
}

@ARTICLE{7081776,
    title="Classification of Disturbances and Cyber-Attacks in Power Systems Using Heterogeneous Time-Synchronized Data",
    author="Shengyi Pan and Morris, T. and Adhikari, U.",
    journal=IEEE_J_IINF,
    month ="Jun.",
    year={2015},
    volume={11},
    number={3},
    pages={650-662},
}

@ARTICLE{stateest,
    title="State estimation for electric power distribution systems in quasi real-time conditions",
    author="I. Roytelman and S. M. Shahidehpour",
    journal=IEEE_J_PWRD,
    month ="Oct.",
    year={1993},
    volume={8},
    number={4},
    pages={2009-2015},
}

@ARTICLE{practical,
    title="Practical aspects of distribution automation in normal and emergency conditions",
    author="I. Roytelman and S. M. Shahidehpour",
    journal=IEEE_J_PWRD,
    month ="Oct.",
    year={1993},
    volume={8},
    number={4},
    pages={2002-2008},
}

@ARTICLE{sign1,
    title="Load Signature Study�Part {I}: Basic Concept, Structure, and Methodology",
    author="J. Liang and Simon Ng and G. Kendall and W. M. John and J. Cheng",
    journal=IEEE_J_PWRD,
    month ="Nov.",
    year={2009},
    volume={25},
    number={2},
    pages={551-560},
}

@ARTICLE{sign2,
    title="Load Signature Study�Part {II}: Disaggregation Framework, Simulation, and Applications",
    author="J. Liang and Simon Ng and G. Kendall and W. M. John and J. Cheng",
    journal=IEEE_J_PWRD,
    month ="Apr.",
    year={2010},
    volume={25},
    number={2},
    pages={561-569},
}

@electronic{newref1,
  title         = "Assessment of Demand Response and Advanced Metering Staff Report",
  url           = "http://www.ferc.gov/legal/staff-reports/2013/oct-demand-response.pdf.",
  organization  = "Federal Energy Regulatory Commission (FERC)",
  month         = "Oct.",
  year          = "2013"
}

@electronic{newref4,
  title         = "Assessment of Demand Response and Advanced Metering Staff Report",
  url           = "http://www.ferc.gov/legal/staff-reports/12-20-12-demand-response.pdf.",
  organization  = "Federal Energy Regulatory Commission (FERC)",
  month         = "Dec.",
  year          = "2012"
}

@electronic{coverageus,
  title         = "Advanced Metering Infrastructure and Customer Systems",
  url           = "http://www.smartgrid.gov/recovery_act/deployment_status/ami_and_customer_systems.",
  organization  = "Recovery Act Smart Grid Programs",
  month         = "Feb.",
  year          = "2014"
}


@electronic{newref2,
  title         = "Smart meter deployments continue to rise",
  url           = "http://www.eia.gov/todayinenergy/detail.cfm?id=8590.",
  organization  = "U.S. Energy Information Administration (EIA)",
  month         = "Nov.",
  year          = "2012"
}

@electronic{degree,
  title         = "Degree of Freedom",
  url           = "http://www.tufts.edu/~gdallal/dof.htm.",
  author        = "Gerard E. Dallal",
  month         = "Mar.",
  year          = "2007"
}

@electronic{regression,
  title="Non-Linear Regression",
  author="Samuel L. Baker",
  url= "http://hspm.sph.sc.edu/Courses/J716/pdf/716-5\%20Non-linear\%20regression.pdf.",
  year= "2006-2008"
}

@electronic{rebuildings,
  title         = "Lean Energy Analysis Using Regression Analysis to Assess Building Energy Performance",
  organization  = "Johnson Controls",
  url           = "http://www.institutebe.com/InstituteBE/media/Library/Resources/Existing\%20Building\%20Retrofits/LEAN-Energy-Analysis_IB.pdf.",
  month         = "Mar.",
  year          = "2013"
}

@ARTICLE{self,
    title="Extraction of Energy Information From Analog Meters Using Image Processing",
    author="Yachen Tang and Chee-Wooi Ten and Chaoli Wang and Gordon Parker",
    journal="Smart Grid, IEEE Transactions on",
    month ="Jan.",
    year="2015",
    volume="6",
    number="4",
    pages="2032-2040",
}

@electronic{linearenergy,
  title         = "Linear Regression Analysis of Energy Consumption Data",
  organization  = "Bizee Software",
  url           = "http://www.degreedays.net/regression-analysis.",
  year          = "2015"
}

@book{least,
author="Robert Jennrich",
title="An Introduction to Computational Statistics",
edition="First",
publisher="Prentice-Hall, Inc.",
year="1995",
pages"4-25"
}

@book{statistics,
author="Thomas Ryan",
title="Modern Regression Methods",
edition="Second",
publisher="John Wiley \& Sons, Inc.",
year="2009",
pages"12-21"
}

@ARTICLE{infrsmgr,
  author        = "Jun-Wei Cao, and Yu-Xin Wan, and Guo-Yu Tu, and Shu-Qing Zhang, and Ai-Xuan Xia, and Xiao-Fei Liu, and Zhen Chen, and Chao Lu, and Ying-Duo Han",
  title         = "Information System Architecture for Smart Grids ",
  journal       = "Chinese Journal of Computers",
  address       = "Beijing, China",
  month         ="Jan.",
  year          ="2013",
  volume        ="36",
  number        ="1",
  pages         ="143--167",
}

@electronic{ftest,
  author        = "Matt Blackwell",
  title         = "Multiple Hypothesis Testing: The F-test",
  url           = "http://www.mattblackwell.org/files/teaching/ftests.pdf.",
  month         = "Dec.",
  year          = "2007"
}





@inproceedings{12467,
  author        = "S. Kumar",
  title         = "Classification and detection of computer intrusions",
  booktitle     = "Ph.D. Dissertation",
  address       = "Department of Computer Science, Purdue University",
  month         = "Aug.",
  year          = "1995"
}

@inproceedings{12519,
  author        = "Y. Xie",
  title         = "A spatiotemporal event correlation approach to computer security",
  booktitle     = "Ph.D. Dissertation (CMU-CS-05-175)",
  address       = "School of Computer Science, Carnegie Mellon University",
  month         = "Aug.",
  year          = "2005"
}

@inproceedings{12571,
  title         = "Vulnerability assessment methodology for electric power infrastructure",
  booktitle     = "US Department of Energy, Office of Energy Assurance",
  month         = "Sep. 30,",
  year          = "2002"
}

@inproceedings{12537,
  author        = "R. K. Fink and D. F. Spencer and R. A. Wells",
  title         = "Lessons learned from cybersecurity assessments of SCADA and energy management systems",
  booktitle     = "US Department of Energy Office of Electricity Delivery and Energy Reliability, National SCADA Test Bed (NSTB)",
  month         = "Sep.",
  year          = "2006"
}

@electronic{12525,
  author        = "M. R. Permann and K. Rohde",
  title         = "Cyber assessment methods for {SCADA} security",
  booktitle     = "US Department of Energy Office of Electricity Delivery and Energy Reliability, Idaho National Laboratory (INL), Jun. 2005",
  url           = "http://www.oe.energy.gov/DocumentsandMedia/Cyber\_Assessment\_Methods \_for\_SCADA\_Security\_Mays\_ISA\_Paper.pdf."
}

@electronic{12521,
  author        = "R. E. Carlson and J. E. Dagle and S. A. Shamsuddin and R. P. Evans",
  title         = "A summary of control system security standards activities in the energy sector",
  howpublished  = "US Department of Energy Office of Electricity Delivery and Energy Reliability, National SCADA Test Bed (NSTB), Oct. 2005",
  url           = "http://www.oe.energy.gov/DocumentsandMedia/Control\_System\_Security \_Standards\_Activities.pdf."
}

@inproceedings{12529,
  author        = "E. Goetz",
  title         = "Cybersecurity of the electric power industry",
  booktitle     = "Report of Investigative Research for Infrastructure Assurance {(IRIA)}",
  address       = "Institute for Security Technology Studies, Dartmouth College",
  month         = "Dec.",
  year          = "2002"
}

@electronic{12551,
  title         = "Information security: Technologies to secure federal systems",
  howpublished  = "{Government Accountability Office {(GAO)} Report to Congressional Requesters}, {GAO-04-467}, Mar. 2004.",
  url           = "http://www.gao.gov/new.items/d04467.pdf.",
}

@electronic{SandiaInv,
  author        = "D. L. King and S. Gonzalez and G. M. Galbraith and W. E. Boyson",
  title         = "Performance Model for Grid-Connected Photovoltaic Inverters",
  howpublished  = "SAND2007-5036",
  url           = "http://infoserve.sandia.gov/sand_doc/2007/075036.pdf",
  year          = "2007"
}


@inproceedings{12515,
  author        = "M. Naedele and D. Dzung and M. Stanimirov",
  title         = "Network security for substation automation systems",
  booktitle     = "Springer-Verlag Berlin, {HeidelbergU. Voges (Ed.): SAFECOMP 2001, LNCS 2187}",
  pages         = "25-34",
  year          = "2001"
}

@inproceedings{12527,
  author        = "P. Baybutt",
  title         = "Cybersecurity risk analysis for process control systems using rings of protection analysis {(ROPA)}",
  booktitle     = "PrimaTech Technical Report, Process Safety Progress",
  volume        = "23",
  number        = "4",
  pages         = "284-290",
  month         = dec,
  year          = "2004"
}

@inproceedings{12393,
  title         = "User manual for the workshop",
  booktitle     = "Cybersecurity standards workshop, {North American Electric Reliability Council (NERC)}",
  address       = "Minneapolis, MN",
  month         = sep,
  year          = "2006"
}

@mastersthesis{NTLPowersys,
  author        = "D. Suriyamongkol",
  title         = "Non-technical losses in electrical power systems",
  school        = "Ohio University",
  address       = "Ohio",
  year          = "2002"
}

the IEEE website
sort key is needed for sorting styles
@electronic{IEEEexample:IEEEwebsite,
  title         = "The {IEEE} Website",
  url           = "http://www.ieee.org/",
  year          = "2007",
  key           = "IEEE"
}


The BibTeX user's guide.
The filename could have been put in the URL instead. But, there might
be other interesting things for the user in the same directory - and
the filename might change before the directory that contains it.
@electronic{IEEEexample:bibtexuser,
  author        = "Oren Patashnik",
  title         = "{{\BibTeX}}ing",
  howpublished  = "{btxdoc.pdf}",
  url           = "http://www.ctan.org/tex-archive/biblio/bibtex/contrib/doc/",
  month         = feb,
  year          = "1988"
}


The BibTeX style designer's guide.
@electronic{IEEEexample:bibtexdesign,
  author        = "Oren Patashnik",
  title         = "Designing {{\BibTeX}} Styles",
  howpublished  = "{btxhak.pdf}",
  url           = "http://www.ctan.org/tex-archive/biblio/bibtex/contrib/doc/",
  month         = feb,
  year          = "1988"
}


A comprehensive BibTeX tutorial.
@electronic{IEEEexample:tamethebeast,
  author        = "Nicolas Markey",
  title         = "Tame the BeaST  ---  The B to X of {{\BibTeX}}",
  url           = "http://tug.ctan.org/tex-archive/info/bibtex/tamethebeast/",
  month         = oct,
  year          = "2005"
}


The BibTeX Tips and FAQ guide.
@electronic{IEEEexample:bibtexFAQ,
  author        = "David Hoadley and Michael Shell",
  title         = "{{\BibTeX}} Tips and {FAQ}",
  howpublished  = "{btxFAQ.pdf}",
  url           = "http://www.ctan.org/tex-archive/biblio/bibtex/contrib/doc/",
  month         = jan,
  year          = "2007"
}


The TeX FAQ
@electronic{IEEEexample:texfaq,
  author        = "Robin Fairbairns",
  title         = "The {{\TeX}} {FAQ}",
  url           = "http://www.tex.ac.uk/cgi-bin/texfaq2html/",
  month         = jan,
  year          = "2007"
}


A BibTeX Guide via Examples.
@electronic{IEEEexample:bibtexguide,
  author        = "Ki-Joo Kim",
  title         = "A {{\BibTeX}} Guide via Examples",
  howpublished  = "{bibtex\_guide.pdf}",
  url           = "http://www.geocities.com/kijoo2000/",
  month         = apr,
  year          = "2004"
}


TeX User Group Bibliography Archive
@electronic{IEEEexample:beebe_archive,
  author        = "Nelson H. F. Beebe",
  title         = "{{\TeX}} User Group Bibliography Archive",
  url           = "http://www.math.utah.edu:8080/pub/tex/bib/index-table.html",
  month         = aug,
  year          = "2006"
}

The natbib.sty package.
@electronic{ctan:natbib,
  author        = "Patrick W. Daly",
  title         = "The natbib.sty package",
  url           = "http://www.ctan.org/tex-archive/macros/latex/contrib/natbib/",
  month         = sep,
  year          = "2006"
}

The url.sty package.
@electronic{IEEEexample:urlsty,
  author        = "Donald Arseneau",
  title         = "The url.sty Package",
  url           = "http://www.ctan.org/tex-archive/macros/latex/contrib/misc/",
  month         = jun,
  year          = "2005",
}


The hyperref.sty package.
@electronic{IEEEexample:hyperrefsty,
  author        = "Sebastian Rahtz and Heiko Oberdiek",
  title         = "The hyperref.sty Package",
  url           = "http://www.ctan.org/tex-archive/macros/latex/contrib/hyperref/",
  month         = nov,
  year          = "2006",
}


The breakurl.sty package.
@electronic{IEEEexample:breakurl,
  author        = "Vilar Camara Neto",
  title         = "The breakurl.sty Package",
  url           = "http://www.ctan.org/tex-archive/macros/latex/contrib/breakurl/",
  month         = aug,
  year          = "2006",
}


The Babel package.
@electronic{IEEEexample:babel,
  author        = "Johannes Braams",
  title         = "The Babel Package",
  url           = "http://www.ctan.org/tex-archive/macros/latex/required/babel/",
  month         = nov,
  year          = "2005",
}


The multibib package.
@electronic{IEEEexample:multibibsty,
  author        = "Thorsten Hansen",
  title         = "The multibib.sty package",
  url           = "http://www.ctan.org/tex-archive/macros/latex/contrib/multibib/",
  month         = jan,
  year          = "2004"
}


The biblatex package.
@electronic{IEEEexample:biblatex,
  author        = "Philipp Lehman",
  title         = "The biblatex package",
  url           = "http://www.ctan.org/tex-archive/macros/latex/exptl/biblatex/",
  month         = jan,
  year          = "2007"
}



The three most common and typical types of references used in
IEEE publications:

an article in a journal
Note the use of the IEEE_J_EDL string, defined in the IEEEabrv.bib file,
for the journal name. IEEEtran.bst defines the BibTeX standard three
letter month codes per IEEE style.
From the June 2002 issue of "IEEE Transactions on Electron Devices",
page 996, reference #16.
@article{IEEEexample:article_typical,
  author        = "S. Zhang and C. Zhu and J. K. O. Sin and P. K. T. Mok",
  title         = "A Novel Ultrathin Elevated Channel Low-temperature
                   Poly-{Si} {TFT}",
  journal       = IEEE_J_EDL,
  volume        = "20",
  month         = nov,
  year          = "1999",
  pages         = "569-571"
}


journal article using et al.
The (five) authors are actually: F. Delorme, S. Slempkes, G. Alibert,
B. Rose, J. Brandon
The month (July) was not given here.
From the September 1998 issue of "IEEE Journal on Selected Areas in
Communications", page 1257, reference #28.
@article{IEEEexample:articleetal,
  author        = "F. Delorme and others",
  title         = "Butt-jointed {DBR} Laser With 15 {nm} Tunability Grown
                   in Three {MOVPE} Steps",
  journal       = "Electron. Lett.",
  volume        = "31",
  number        = "15",
  year          = "1995",
  pages         = "1244-1245"
}


a paper in a conference proceedings
"conference" can be used as an alias for "inproceedings"
From the June 2002 issue of "Journal of Microelectromechanical Systems",
page 205, reference #16.
@inproceedings{IEEEexample:conf_typical,
  author        = "R. K. Gupta and S. D. Senturia",
  title         = "Pull-in Time Dynamics as a Measure of Absolute Pressure",
  booktitle     = "Proc. {IEEE} International Workshop on
                   Microelectromechanical Systems ({MEMS}'97)",
  address       = "Nagoya, Japan",
  month         = jan,
  year          = "1997",
  pages         = "290-294"
}


a book
From the May 2002 issue of "IEEE Transactions on Magnetics",
page 1466, reference #4.
@book{IEEEexample:book_typical,
  author        = "B. D. Cullity",
  title         = "Introduction to Magnetic Materials",
  publisher     = "Addison-Wesley",
  address       = "Reading, MA",
  year          = "1972"
}




Other examples

journal article with large page numbers, IEEE will divide numbers 5 digits
or longer into groups of three with small spaces between them. Page ranges
can be separated via either "-" or "--", IEEEtran.bst will automatically
convert the separator dash(es) to "--".
Authors and/or IEEE do not always provide/use the journal number, but it
was used in this case. IEEEtran.bst can be configured to ignore journal
numbers if desired.
From the August 2000 issue of "IEEE Photonics Technology Letters",
page 1039, reference #11.
@article{IEEEexample:articlelargepages,
  author        = "A. Castaldini and A. Cavallini and B. Fraboni
                   and P. Fernandez and J. Piqueras",
  title         = "Midgap Traps Related to Compensation Processes in
                   {CdTe} Alloys",
  journal       = "Phys. Rev. B.",
  volume        = "56",
  number        = "23",
  year          = "1997",
  pages         = "14897-14900"
}


journal article with dual months
use the BibTeX "#" concatenation operator
From the March 2002 issue of "IEEE Transactions on Mechatronics",
page 21, reference #8.
@article{IEEEexample:articledualmonths,
  author        = "Y. Okada and K. Dejima and T. Ohishi",
  title         = "Analysis and Comparison of {PM} Synchronous Motor and
                   Induction Motor Type Magnetic Bearings",
  journal       = IEEE_J_IA,
  volume        = "31",
  month         = sep # "/" # oct,
  year          = "1995",
  pages         = "1047-1053"
}


journal article to be published as a misc entry type
date information like month and year is optional
However, the article form like that below may be a better approach.
From the May 2002 issue of "IEEE Journal of Selected Areas in
Communication", page 725, reference #3.
@misc{IEEEexample:TBPmisc,
  author        = "M. Coates and A. Hero and R. Nowak and B. Yu",
  title         = "Internet Tomography",
  howpublished  = IEEE_M_SP,
  month         = may,
  year          = "2002",
  note          = "to be published"
}


journal article to be published as an article entry type
year is required, so if absent, use the year field to hold
the "submitted for publication" in order to avoid a warning for
the missing year field.
From the June 2002 issue of "IEEE Transactions on Information Theory",
page 1461, reference #21.
@article{IEEEexample:TBParticle,
  author        = "N. Kahale and R. Urbanke",
  title         = "On the Minimum Distance of Parallel and Serially
                   Concatenated Codes",
  journal       = IEEE_J_IT,
  year          = "submitted for publication"
}

book with editor and no author
From the June 2002 issue of "IEEE Transactions on Information Theory",
page 1725, reference #1.
@book{IEEEexample:bookwitheditor,
  editor        = "J. C. Candy and G. C. Temes",
  title         = "Oversampling Delta-Sigma Data Converters Theory,
                   Design and Simulation",
  publisher     = "{IEEE} Press.",
  address       = "New York",
  year          = "1992"
}


book with edition, author and editor
Note that the standard BibTeX styles do not support book entries with both
author and editor fields, but IEEEtran.bst does.
The standard BibTeX way of specifying the edition is to use capitalized
ordinal words such as "First", "Second", etc. Most .bst files can convert
up to about "Fifth" into the format needed. IEEEtran.bst can convert up
to "Tenth" to the "Arabic ordinal" form IEEE uses (e.g., "10th"). For
editions over the tenth, it is best to use the "Arabic ordinal" form for
IEEE related work (e.g., "101st")
Note how "Jr." has to be entered.
From the May 2002 issue of "Journal of Lightwave Technology", page 856,
reference #17.
@book{IEEEexample:book,
  author        = "S. M. Metev and V. P. Veiko",
  editor        = "Osgood, Jr., R. M.",
  title         = "Laser Assisted Microtechnology",
  edition       = "Second",
  publisher     = "Springer-Verlag",
  address       = "Berlin, Germany",
  year          = "1998"
}


book with series and volume
From the January 2000 issue of "IEEE Transactions on Communications",
page 11, reference #31.
@book{IEEEexample:bookwithseriesvolume,
  editor        = "J. Breckling",
  title         = "The Analysis of Directional Time Series: Applications to
                   Wind Speed and Direction",
  series        = "Lecture Notes in Statistics",
  publisher     = "Springer",
  address       = "Berlin, Germany",
  year          = "1989",
  volume        = "61"
}


inbook with chapter number. The pages field could also have been given.
The chapter number could be changed to something else such as a section
number via the type field: type = "sec.".
From the May 2002 issue of "IEEE Transactions on Circuits and Systems---I:
Fundamental Applications and Theory", page 638, reference #22.
@inbook{IEEEexample:inbook,
  author        = "H. E. Rose",
  title         = "A Course in Number Theory",
  publisher     = "Oxford Univ. Press",
  address       = "New York, NY",
  year          = "1988",
  chapter       = "3"
}


inbook with pages and note. The language field is not set to Russian
because the title is presented here in its translated, English, form.
From the May 2002 issue of "IEEE Transactions on Magnetics", page 1533,
reference #5.
@inbook{IEEEexample:inbookpagesnote,
  author        = "B. K. Bul",
  title         = "Theory Principles and Design of Magnetic Circuits",
  publisher     = "Energia Press",
  address       = "Moscow",
  year          = "1964",
  pages         = "464",
  note          = "(in Russian)"
}





incollection with author and editor
From the May 2002 issue of "Journal of Lightwave Technology",
page 807, reference #7.
@incollection{IEEEexample:incollection,
  author        = "W. V. Sorin",
  editor        = "D. Derickson",
  title         = "Optical Reflectometry for Component Characterization",
  booktitle     = "Fiber Optic Test and Measurement",
  publisher     = "Prentice-Hall",
  address       = "Englewood Cliffs, NJ",
  year          = "1998"
}


incollection with series
From the April 2000 issue of "IEEE Transactions on Communication",
page 609, reference #3.
@incollection{IEEEexample:incollectionwithseries,
  author        = "J. B. Anderson and K. Tepe",
  title         = "Properties of the Tailbiting {BCJR} Decoder",
  booktitle     = "Codes, Systems and Graphical Models",
  series        = "{IMA} Volumes in Mathematics and Its Applications",
  publisher     = "Springer-Verlag",
  address       = "New York",
  year          = "2000"

}


incollection with author, editor, chapter and pages
From the January 2000 issue of "IEEE Transactions on Communications",
page 16, reference #9.
@incollection{IEEEexample:incollection_chpp,
  author        = "P. Hedelin and P. Knagenhjelm and M. Skoglund",
  editor        = "W. B. Kleijn and K. K. Paliwal",
  title         = "Theory for Transmission of Vector Quantization Data",
  booktitle     = "Speech Coding and Synthesis",
  publisher     = "Elsevier Science",
  address       = "Amsterdam, The Netherlands",
  year          = "1995",
  chapter       = "10",
  pages         = "347-396"
}


incollection with a large number of authors, some authors/journals will
use et al. for so many names. IEEEtran.bst can be configured to do this
if desired, or "R. M. A. Dawson and others" can be used instead.
Note that IEEE may not include the publisher for incollection entries -
IEEEtran.bst will not issue a warning if the publisher is missing for
incollections - but other .bst files often will.
From the June 2002 issue of "IEEE Transactions on Electron Devices",
page 996, reference #12.
@incollection{IEEEexample:incollectionmanyauthors,
  author        = "R. M. A. Dawson and Z. Shen and D. A. Furst and
                   S. Connor and J. Hsu and M. G. Kane and R. G. Stewart and
                   A. Ipri and C. N. King and P. J. Green and R. T. Flegal
                   and S. Pearson and W. A. Barrow and E. Dickey and K. Ping
                   and C. W. Tang and S. Van. Slyke and
                   F. Chen and J. Shi and J. C. Sturm and M. H. Lu",
  title         = "Design of an Improved Pixel for a Polysilicon
                   Active-Matrix Organic {LED} Display",
  booktitle     = "{SID} Tech. Dig.",
  volume        = "29",
  year          = "1998",
  pages         = "11-14"
}





A Motorola data book as a manual
It is somewhat unusual to include the data book part number.
in the title. It might be more correct to put this information
in the howpublished field instead.
From the December 2000 issue of "IEEE Transactions on Communications",
page 1986, reference #10.
@manual{IEEEexample:motmanual,
  title         = "{FLEXChip} Signal Processor ({MC68175/D})",
  organization  = "Motorola",
  year          = "1996"
}

@manual{manual:ANSI,
  title         = "ANSI C12.20-2002 - Electricity Meters 0.2 and 0.5 Accuracy Classes",
  organization  = "ANSI",
  year          = "2002"
}

@Book{Book:WHKersting,
 author    = "W.~H. Kersting",
 title     = "Distribution System Modeling and Analysis",
 publisher = "CRC Press.",
 year      =  2002,
 edition   = "2nd Edition"
}

same reference, but using IEEEtran's howpublished extension
@manual{IEEEexample:motmanualhowpub,
  title         = "{FLEXChip} Signal Processor",
  howpublished  = "{MC68175/D}",
  organization  = "Motorola",
  year          = "1996"
}




conference paper with an address and days. Some journals capitalize the
letters in "Globecom", this one did not.
From the May 2002 issue of "IEEE Transactions on Communications",
page 697, reference #12.
@inproceedings{IEEEexample:confwithadddays,
  author        = "M. S. Yee and L. Hanzo",
  title         = "Radial Basis Function Decision Feedback Equaliser
                   Assisted Burst-by-burst Adaptive Modulation",
  booktitle     = "Proc. {IEEE} Globecom '99",
  address       = "Rio de Janeiro, Brazil",
  month         = dec # " 5--9,",
  year          = "1999",
  pages         = "2183-2187"
}


conference paper with volume number. A conference proceedings with a vol
number is a little uncommon, note that the vol number is placed
before the address in the formatted bibliography entry
From the April 2002 issue of "IEEE/ACM Transactions on Networking",
page 181, reference #26.
@inproceedings{IEEEexample:confwithvolume,
  author        = "M. Yajnik and S. B. Moon and J. Kurose and D. Towsley",
  title         = "Measurement and Modeling of the Temporal Dependence in
                   Packet Loss",
  booktitle     = "Proc. {IEEE} {INFOCOM}'99",
  volume        = "1",
  address       = "New York, NY",
  month         = mar,
  year          = "1999",
  pages         = "345-352"
}


conference paper with a paper number, the type field can be used to
override the word "paper", e.g., type = "postdeadline paper". A type
can be given even without a paper  field.
Also, IEEEtran.bst can be configured to ignore paper numbers and types.
From the May 2002 issue of "Journal of Lightwave Technology",
page 807, reference #4.
@inproceedings{IEEEexample:confwithpaper,
  author        = "M. Wegmuller and J. P. von der Weid and P. Oberson
                   and N. Gisin",
  title         = "High Resolution Fiber Distributed Measurements With
                   Coherent {OFDR}",
  booktitle     = "Proc. {ECOC}'00",
  year          = "2000",
  paper         = "11.3.4",
  pages         = "109"
}


conference paper with a postdeadline type paper, the paper field is
optional.
From the August 2000 issue of "IEEE Photonics Technology Letters",
page 1087, reference #12.
@inproceedings{IEEEexample:confwithpapertype,
  author        = "B. Mikkelsen and G. Raybon and R.-J. Essiambre and
                   K. Dreyer and Y. Su. and L. E. Nelson and J. E. Johnson
                   and G. Shtengel and A. Bond and D. G. Moodie and
                   A. D. Ellis",
  title         = "160 {Gbit/s} Single-channel Transmission Over 300 km
                   Nonzero-dispersion Fiber With Semiconductor Based
                   Transmitter and Demultiplexer",
  booktitle     = "Proc. {ECOC}'99",
  year          = "1999",
  paper         = "2-3",
  type          = "postdeadline paper",
  pages         = "28-29"
}


presented at a conference
intype overrides the default "in" and causes the booktitle not to be
emphasized (rendered in italics).
From the February 2002 issue of "IEEE/ACM Transactions on Networking",
page 163, reference #6.
@inproceedings{IEEEexample:presentedatconf,
  author        = "S. G. Finn and M. M{\'e}dard and R. A. Barry",
  title         = "A Novel Approach to Automatic Protection Switching
                   Using Trees",
  intype        = "presented at the",
  booktitle     = "Proc. Int. Conf. Commun.",
  year          = "1997"
}





master's thesis, often the University name will be abbreviated and the
state or country will be included in the address. The type field can
used to override the default type "Master's thesis"
From the June 2002 issue of "IEEE Transactions on Microelectromechanical
Systems", page 186, reference #11.
@mastersthesis{IEEEexample:masters,
  author        = "Nin C. Loh",
  title         = "High-Resolution Micromachined Interferometric
                   Accelerometer",
  school        = "Massachusetts Institute of Technology",
  address       = "Cambridge",
  year          = "1992"
}


master's thesis with a type field
From the August 2001 issue of "IEEE/ACM Transactions on Networking",
page 391, reference #12.
@mastersthesis{IEEEexample:masterstype,
  author        = "A. Karnik",
  title         = "Performance of {TCP} Congestion Control with Rate
                   Feedback: {TCP/ABR} and Rate Adaptive {TCP/IP}",
  school        = "Indian Institute of Science",
  type          = "M. Eng. thesis",
  address       = "Bangalore, India",
  month         = jan,
  year          = "1999"
}





Ph.D. dissertation with a URL field, the university is abbreviated
From the October 2001 issue of "IEEE/ACM Transactions on Networking",
page 590, reference #11.
@phdthesis{IEEEexample:phdurl,
  author        = "Q. Li",
  title         = "Delay Characterization and Performance Control of
                   Wide-area Networks",
  school        = "Univ. of Delaware",
  address       = "Newark",
  month         = may,
  year          = "2000",
  url           = "http://www.ece.udel.edu/~qli"
}





technical report
From the August 2001 issue of "IEEE/ACM Transactions on Networking",
page 490, reference #15.
@techreport{IEEEexample:techrep,
  author        = "R. Jain and K. K. Ramakrishnan and D. M. Chiu",
  title         = "Congestion Avoidance in Computer Networks with a
                   Connectionless Network Layer",
  institution   = "Digital Equipment Corporation",
  address       = "MA",
  number        = "DEC-TR-506",
  month         = aug,
  year          = "1987"
}


technical report with type
for those times when "Tech. Rep." needs to be modified
From the February 2001 issue of "IEEE/ACM Transactions on Networking",
page 46, reference #8.
@techreport{IEEEexample:techreptype,
  author        = "J. Padhye and V. Firoiu and D. Towsley",
  title         = "A Stochastic Model of {TCP} {R}eno Congestion Avoidance
                   and Control",
  institution   = "Univ. of Massachusetts",
  address       = "Amherst, MA",
  type          = "CMPSCI Tech. Rep.",
  number        = "99-02",
  year          = "1999"
}


technical report with type
for those times when "Tech. Rep." needs to be modified
This reference did not have an address.
From the January 2000 issue of "IEEE Transactions on Communications",
page 117, reference #6.
@techreport{IEEEexample:techreptypeii,
  author        = "D. Middleton and A. D. Spaulding",
  title         = "A Tutorial Review of Elements of Weak Signal Detection
                   in Non-{G}aussian {EMI} Environments",
  institution   = "National Telecommunications and Information
                   Administration ({NTIA}), U.S. Dept. of Commerce",
  type          = "NTIA Report",
  number        = "86-194",
  month         = may,
  year          = "1986"
}





an unpublished work
for unpublished types, the note field is required. IEEE usually
just uses the word "unpublished" for the note.
From the August 2001 issue of "IEEE/ACM Transactions on Networking",
page 391, reference #16.
@unpublished{IEEEexample:unpublished,
  author        = "T. J. Ott and N. Aggarwal",
  title         = "{TCP} over {ATM}: {ABR} or {UBR}",
  note          = "Unpublished"
}





electronic with a howpublished information field
From the August 2001 issue of "IEEE/ACM Transactions on Networking",
page 391, reference #7.
@electronic{IEEEexample:electronhowinfo,
  author        = "V. Jacobson",
  title         = "Modified {TCP} Congestion Avoidance Algorithm",
  howpublished  = "end2end-interest mailing list",
  url           = "ftp://ftp.isi.edu/end2end/end2end-interest-1990.mail",
  month         = apr,
  year          = "1990"
}


electronic with a howpublished information field
From the August 2001 issue of "IEEE/ACM Transactions on Networking",
page 418, reference #31.
@electronic{IEEEexample:electronhowinfo2,
  author        = "V. Valloppillil and K. W. Ross",
  title         = "Cache Array Routing Protocol v1.1",
  howpublished  = "Internet draft",
  url           = "http://ds1.internic.net/internet-drafts/draft-vinod-carp-v1-03.txt",
  year          = "1998"
}


electronic with an organization and address
From the February 2002 issue of "IEEE/ACM Transactions on Networking",
page 114, reference #15.
@electronic{IEEEexample:electronorgadd,
  author        = "D. H. Lorenz and A. Orda",
  title         = "Optimal Partition of {QoS} Requirements on Unicast
                   Paths and Multicast Trees",
  organization  = "Dept. Elect. Eng., Technion",
  address       = "Haifa, Israel",
  url           = "ftp://ftp.technion.ac.il/pub/supported/ee/Network/lor.mopq98.ps",
  month         = jul,
  year          = "1998"
}





U.S. patent
Use the type field to override the patent type. e.g.,
type = "Patent Application"
The address is that of the assignee. Note that IEEE does not
display the assignee, the address, and only displays one date.
(if year is not present, the filed dates are used.) However, this
information should be entered as other BibTeX styles may use it.
alternatively, nationality could be entered as "U.S."
From the April 2000 issue of "IEEE Transactions on Communications",
page 542, reference #6.
@patent{IEEEexample:uspat,
  author        = "Ronald E. Sorace and Victor S. Reinhardt and
                   Steven A. Vaughn",
  assignee      = "Hughes Aircraft Company",
  address       = "Los Angeles, CA",
  title         = "High-Speed Digital-to-{RF} Converter",
  nationality   = "United States",
  number        = "5668842",
  dayfiled      = "28",
  monthfiled    = feb,
  yearfiled     = "1995",
  day           = "16",
  month         = sep,
  year          = "1997"
}


Japanese Patent
From the April 2000 issue of "IEEE Transactions on Communications",
page 556, reference #6.
@patent{IEEEexample:jppat,
  author        = "U. Hideki",
  title         = "Quadrature Modulation Circuit",
  nationality   = "Japanese",
  number        = "152932/92",
  day           = "20",
  month         = may,
  year          = "1992"
}


French Patent request, the language field must be entered in lower case
as this is the option name Babel uses. The nationality field needs to be
capitalized. Because this is a patent request, the date filed fields are
used while the date fields are left empty/missing. In other countries,
the words "Patent Application", etc. are used instead.
From the April 2000 issue of "IEEE Transactions on Communications",
page 556, reference #9.
@patent{IEEEexample:frenchpatreq,
  author        = "F. Kowalik and M. Isard",
  title         = "Estimateur d'un D{\'e}faut de Fonctionnement
                   d'un Modulateur en Quadrature et {\'E}tage de Modulation
                   l'Utilisant",
  language      = "french",
  nationality   = "French",
  type          = "Patent Request",
  number        = "9500261",
  dayfiled      = "11",
  monthfiled    = jan,
  yearfiled     = "1995"
}





a periodical
From the April 2001 issue of "IEEE/ACM Transactions on Networking",
page 160, reference #1.
sort key is needed for sorting styles
@periodical{IEEEexample:periodical,
  title         = IEEE_M_PCOM # ", Special Issue on Wireless {ATM}",
  volume        = "3",
  month         = aug,
  year          = "1996",
  key           = "IEEE"
}





standard, IEEE does not use the address for standards, but it is good
to provide one for BibTeX styles that use it.
From the August 2001 issue of "IEEE/ACM Transactions on Networking",
page 451, reference #2.
@standard{IEEEexample:standard,
  title         = "Wireless {LAN} Medium Access Control {(MAC)} and
                   Physical Layer {(PHY)} Specification",
  organization  = "IEEE",
  address       = "Piscataway, NJ",
  number        = "802.11",
  year          = "1997"
}


standard with type and revision, the type overrides the word standard
(or std.). Here a standard number is not available and a revision number
is used.
From the August 2000 issue of "IEEE Photonics Technology Letters",
page 1048, reference #13.
@standard{IEEEexample:standardproposed,
  title         = "Fiber Channel Physical Interface ({FC-PI})",
  institution   = "NCITS",
  address       = "Washington, DC",
  type          = "Working Draft Proposed Standard",
  revision      = "5.2",
  year          = "1999"
}


standard draft as a misc with author
From the May 2002 issue of "IEEE Journal of Selected Areas in
Communication", page 725, reference #16.
@misc{IEEEexample:draftasmisc,
  author        = "I. Widjaja and A. Elwalid",
  title         = "{MATE}: {MPLS} Adaptive Traffic Engineering",
  howpublished  = "IETF Draft",
  year          = "1999"
}





misc for a techreport like reference
techreport is not perfectly suitable because this entry lacks
an institution field
From the August 2001 issue of "IEEE/ACM Transactions on Networking",
page 490, reference #22.
@misc{IEEEexample:miscforum,
  author        = "L. Roberts",
  title         = "Enhanced Proportional Rate Control Algorithm {PRCA}",
  howpublished  = "{ATM} Forum Contribution 94-0735R1",
  month         = aug,
  year          = "1994"
}


misc for a white paper
From the August 2001 issue of "IEEE/ACM Transactions on Networking",
page 478, reference #4 - Note that the reference there (improperly?)
used the author field for "Cisco".
@misc{IEEEexample:whitepaper,
  title         = "Advanced {QoS} Services for the Intelligent Internet",
  howpublished  = "White Paper",
  organization  = "Cisco",
  month         = may,
  year          = "1997"
}


misc for a data sheet
From the November 2000 issue of "IEEE Photonics Technology Letters",
page 1551, reference #6.
@misc{IEEEexample:datasheet,
  title         = "{PDCA12-70} Data Sheet",
  organization  = "Opto Speed SA",
  address       = "Mezzovico, Switzerland"
}





Other unusual references

a private communication as a misc entry type
sometimes the designation "personal communication" is used instead
From the June 2002 issue of "IEEE Transactions on Information Theory",
page 1725, reference #16.
@misc{IEEEexample:private,
  author        = "S. Konyagin",
  howpublished  = "private communication",
  year          = "1998"
}


an internet request for comments (RFC) as a misc entry type
It would also be nice to add a URL to these types of things.
RFCs can also be handled as electronic references.
From the April 2002 issue of "IEEE/ACM Transactions on Networking",
page 181, reference #14.
@misc{IEEEexample:miscrfc,
  author        = "K. K. Ramakrishnan and S. Floyd",
  title         = "A Proposal to Add Explicit Congestion
                   Notification ({ECN}) to {IP}",
  howpublished  = "RFC 2481",
  month         = jan,
  year          = "1999"
}


a software package as a manual
From the June 2002 issue of "IEEE/ASME Journal of Microelectromechanical
Systems", page 205, reference #20.
Sometimes they put the version/release information in the title.
@manual{IEEEexample:softmanual,
  title         = "SaberDesigner Reference Manual",
  organization  = "Analogy, Inc.",
  address       = "Beaverton, OR",
  year          = "1998",
  note          = "Release 4.3"
}


a software package as an electronic reference
From the February 2003 issue of  "IEEE/ACM Transactions on Networking",
page 46, reference #24. If there is no author or organization, a sorting
key is required for sorting styles. It might be a good idea to include
month and year fields as well.
@electronic{IEEEexample:softonline,
  title         = "Ucb/lbnl/vint Network Simulator---ns (Version 2)",
  url           = "http://www-mash.cs.berkeley.edu/ns/",
  key           = "ns"
}


misc for a German regulation
In German, the first letters of nouns are capitalized, so we do so here.
From the June 2002 issue of "IEEE Journal in Selected Areas in
Communication", page 892, reference #9.
@misc{IEEEexample:miscgermanreg,
  title         = "{M}essung von {S}t{\"o}rfeldern an {A}nlagen
                   und {L}eitungen der {T}elekommunikation im
                   {F}requenzbereich 9 {kHz} bis 3 {GHz}",
  language      = "german",
  howpublished  = "{M}e{\ss}vorschrift {R}eg {TP} {MV} 05",
  organization  = "Regulierungsbeh{\"o}rde f{\"u}r {T}elekommunikation und
                   {P}ost ({R}eg {TP})"
}



Ways to handle things like CCSDS's Blue Books
journal article with a URL. However, this is not a very good approach
because the Blue Books are not really journals and the author field has
to be abused.
From the June 2002 issue of "IEEE Transactions on Information Theory",
page 1461, reference #7.
@article{IEEEexample:bluebookarticle,
  author        = "{Consulative Committee for Space Data Systems (CCSDS)}",
  title         = "Telemetry Channel Coding",
  journal       = "Blue Book",
  number        = "4",
  year          = "1999",
  url           = "http://www.ccsds.org/documents/pdf/CCSDS-101.0-B-4.pdf"
}


CCSDS's Blue Book handled as a book
However, it is not a good idea to have to use the author field for
an organization (done here because the book entry type requires an
author field).
@book{IEEEexample:bluebookbook,
  author        = "{Consulative Committee for Space Data Systems (CCSDS)}",
  title         = "Telemetry Channel Coding",
  series        = "Blue Book",
  number        = "4",
  publisher     = "{CCSDS}",
  address       = "Newport Beach, {CA}",
  year          = "1999",
  url           = "http://www.ccsds.org/documents/pdf/CCSDS-101.0-B-4.pdf"
}


CCSDS's Blue Book handled as a manual
This is a much better approach, but uses the howpublished field.
@manual{IEEEexample:bluebookmanual,
  title         = "Telemetry Channel Coding",
  howpublished  = "ser. Blue Book, No. 4",
  organization  = "Consulative Committee for Space Data Systems (CCSDS)",
  address       = "Newport Beach, CA",
  year          = "1999",
  url           = "http://www.ccsds.org/documents/pdf/CCSDS-101.0-B-4.pdf"
}





CCSDS's Blue Book handled as a standard
Probably the best approach for this particular case because the work
is standard related. Note that IEEE does not display the address for
standards.
@standard{IEEEexample:bluebookstandard,
  title         = "Telemetry Channel Coding",
  howpublished  = "ser. Blue Book, No. 4",
  organization  = "Consulative Committee for Space Data Systems (CCSDS)",
  address       = "Newport Beach, CA",
  type          = "Recommendation for Space Data System Standard",
  number        = "101.0-B-4",
  month         = May,
  year          = "1999",
  url           = "http://www.ccsds.org/documents/pdf/CCSDS-101.0-B-4.pdf"
}

@book{UGknight,
  author        = "U. G. Knight",
  title         = "Power Systems in Emergencies: From Contingency Planning to Crisis Management",
  edition       = "1",
  publisher     = "Wiley",
  address       = "Baffins Lane, Chichester, {UK}",
  year          = "2001",
  url           = ""
}

@misc{TB432,
author={\relax{CIGRE Working Group B5.19}},
title={\relax{Protection Relay Coordination}},
howpublished={Technical Brochure 432},
year={2010},
month={Oct.},
pages={1-179},
}


@techreport{TB566,
  author        = "Working Group C4.605",
  title         = "Modelling and Aggregation of Loads
in Flexible Power Networks",
  institution   = "CIGRE",
  address       = "Paris",
  number        = "TB566",
  month         = "Feb.",
  year          = "2014"
}

@article{9078877,
  author        = "Koji Yamashita and Chee-Wooi Ten and Yeonwoo Rho and Lingfeng Wang and Wei Wei and Andrew Francis Ginter",
  title         = "Measuring Systemic Risk of Switching Attacks Based on Cybersecurity Technologies in Substations",
  journal       = "IEEE Transactions on Power Systems",
  number        = "",
  year          = "2020",
  pubstate      = "Early access",
}





An example of a IEEEtran control entry which can change some IEEEtran.bst
settings. An entry like this must be cited via \bstctlcite{} command
before the first real \cite{}. The same entry key cannot be called twice
(just like multiple \cite{} of the same entry key place only one entry
in the bibliography.)
The available control fields are:

CTLuse_article_number
"no" turns off the display of the number for articles.
"yes" enables

CTLuse_paper
"no" turns off the display of the paper and type fields in inproceedings.
"yes" enables

CTLuse_forced_etal
"no" turns off the forced use of "et al."
"yes" enables

CTLmax_names_forced_etal
The maximum number of names that can be present beyond which an "et al."
usage is forced. Be sure that CTLnames_show_etal (below)
is not greater than this value!

CTLnames_show_etal
The number of names that will be shown with a forced "et al.".
Must be less than or equal to CTLmax_names_forced_etal

CTLuse_alt_spacing
"no" turns off the alternate interword spacing for entries with URLs.
"yes" enables

CTLalt_stretch_factor
If alternate interword spacing for entries with URLs is enabled, this is
the interword spacing stretch factor that will be used. For example, the
default "4" here means that the interword spacing in entries with URLs can
stretch to four times normal. Does not have to be an integer.

CTLdash_repeated_names
"no" turns off the "dashification" of repeated (i.e., identical to those
of the previous entry) names. IEEE normally does this.
"yes" enables

CTLname_format_string
The name format control string as explained in the BibTeX style hacking
guide.
IEEE style "{f.~}{vv~}{ll}{, jj}" is the default,

CTLname_latex_cmd
A LaTeX command that each name will be fed to (e.g., "\textsc").
Leave empty if no special font is desired for the names.
The default is empty.

CTLname_url_prefix
The prefix text used before URLs.
The default is "[Online]. Available:" A space will be inserted after this
text. If this space is not wanted, just use \relax at the end of the
prefix text.


Those fields that are not to be changed can be left out.
@IEEEtranBSTCTL{IEEEexample:BSTcontrol,
  CTLuse_article_number     = "yes",
  CTLuse_paper              = "yes",
  CTLuse_forced_etal        = "no",
  CTLmax_names_forced_etal  = "10",
  CTLnames_show_etal        = "1",
  CTLuse_alt_spacing        = "yes",
  CTLalt_stretch_factor     = "4",
  CTLdash_repeated_names    = "yes",
  CTLname_format_string     = "{f.~}{vv~}{ll}{, jj}",
  CTLname_latex_cmd         = "",
  CTLname_url_prefix        = "[Online]. Available:"
}


\begin{thebibliography}{10}
\providecommand{\url}[1]{#1}
\csname url@samestyle\endcsname
\providecommand{\newblock}{\relax}
\providecommand{\bibinfo}[2]{#2}
\providecommand{\BIBentrySTDinterwordspacing}{\spaceskip=0pt\relax}
\providecommand{\BIBentryALTinterwordstretchfactor}{4}
\providecommand{\BIBentryALTinterwordspacing}{\spaceskip=\fontdimen2\font plus
\BIBentryALTinterwordstretchfactor\fontdimen3\font minus
  \fontdimen4\font\relax}
\providecommand{\BIBforeignlanguage}[2]{{%
\expandafter\ifx\csname l@#1\endcsname\relax
\typeout{** WARNING: IEEEtran.bst: No hyphenation pattern has been}%
\typeout{** loaded for the language `#1'. Using the pattern for}%
\typeout{** the default language instead.}%
\else
\language=\csname l@#1\endcsname
\fi
#2}}
\providecommand{\BIBdecl}{\relax}
\BIBdecl

\bibitem{9215386}
K.~Poornesh, K.~P. Nivya, and K.~Sireesha, ``A comparative study on electric
  vehicle and internal combustion engine vehicles,'' in \emph{2020
  International Conference on Smart Electronics and Communication (ICOSEC)},
  2020, pp. 1179--1183.

\bibitem{9786788}
A.~Ahmad, Z.~Qin, T.~Wijekoon, and P.~Bauer, ``An overview on medium voltage
  grid integration of ultra-fast charging stations: Current status and future
  trends,'' \emph{IEEE Open Journal of the Industrial Electronics Society},
  vol.~3, pp. 420--447, 2022.

\bibitem{9336258}
L.~Wang, Z.~Qin, T.~Slangen, P.~Bauer, and T.~van Wijk, ``Grid impact of
  electric vehicle fast charging stations: Trends, standards, issues and
  mitigation measures - an overview,'' \emph{IEEE Open Journal of Power
  Electronics}, vol.~2, pp. 56--74, 2021.

\bibitem{9272723}
S.~Acharya, Y.~Dvorkin, H.~Pandžić, and R.~Karri, ``Cybersecurity of smart
  electric vehicle charging: A power grid perspective,'' \emph{IEEE Access},
  vol.~8, pp. 214\,434--214\,453, 2020.

\bibitem{9369743}
M.~Girdhar, J.~Hong, R.~Karnati, S.~Lee, and S.~Choi, ``Cybersecurity of
  process bus network in digital substations,'' in \emph{2021 International
  Conference on Electronics, Information, and Communication (ICEIC)}, 2021, pp.
  1--6.

\bibitem{7785783}
A.~Valdes, R.~Macwan, and M.~Backes, ``Anomaly detection in electrical
  substation circuits via unsupervised machine learning,'' in \emph{2016 IEEE
  17th International Conference on Information Reuse and Integration (IRI)},
  2016, pp. 500--505.

\bibitem{9916758}
J.~Hong, M.~Girdhar, C.-W. Ten, S.~Lee, and S.~Choi, ``Cybersecurity of sampled
  value messages in substation automation system,'' in \emph{2022 IEEE Power \&
  Energy Society General Meeting (PESGM)}, 2022, pp. 1--1.

\bibitem{8493213}
M.~Mültin, ``Iso 15118 as the enabler of vehicle-to-grid applications,'' in
  \emph{2018 International Conference of Electrical and Electronic Technologies
  for Automotive}, 2018, pp. 1--6.

\bibitem{905746}
T.~Krause, R.~Ernst, B.~Klaer, I.~Hacker, and M.~Henze, ``Cybersecurity in
  power grids: Challenges and opportunities,'' \emph{arXiv}, 2021.

\bibitem{7807218}
D.~T. {Hoang}, P.~{Wang}, D.~{Niyato}, and E.~{Hossain}, ``Charging and
  discharging of plug-in electric vehicles (pevs) in vehicle-to-grid (v2g)
  systems: A cyber insurance-based model,'' \emph{IEEE Access}, vol.~5, pp.
  732--754, 2017.

\bibitem{9583592}
M.~Girdhar, J.~Hong, H.~Lee, and T.-J. Song, ``Hidden markov models-based
  anomaly correlations for the cyber-physical security of ev charging
  stations,'' \emph{IEEE Transactions on Smart Grid}, vol.~13, no.~5, pp.
  3903--3914, 2022.

\bibitem{9490069}
A.~Sanghvi and T.~Markel, ``Cybersecurity for electric vehicle fast-charging
  infrastructure,'' in \emph{2021 IEEE Transportation Electrification
  Conference \& Expo (ITEC)}, 2021, pp. 573--576.

\bibitem{8030362}
A.~{Bindra}, ``Securing the power grid: Protecting smart grids and connected
  power systems from cyberattacks,'' \emph{IEEE Power Electronics Magazine},
  vol.~4, no.~3, pp. 20--27, 2017.

\bibitem{9698889}
A.~Hafeez, J.~Mohan, M.~Girdhar, and S.~S. Awad, ``Machine learning based ecu
  detection for automotive security,'' in \emph{2021 17th International
  Computer Engineering Conference (ICENCO)}, 2021, pp. 73--81.

\bibitem{905fg578}
R.~I. Y.~Yusoff and Z.~Hassan, ``Common phases of computer forensics
  investigation models,'' \emph{International Journal of Computer Science \&
  Information Technology (IJCSIT)}, vol.~3, no.~3, 2011.

\bibitem{KORONIOTIS202091}
\BIBentryALTinterwordspacing
N.~Koroniotis, N.~Moustafa, and E.~Sitnikova, ``A new network forensic
  framework based on deep learning for internet of things networks: A particle
  deep framework,'' \emph{Future Generation Computer Systems}, vol. 110, pp.
  91--106, 2020. [Online]. Available:
  \url{https://www.sciencedirect.com/science/article/pii/S0167739X19325105}
\BIBentrySTDinterwordspacing

\bibitem{9243536}
H.~I. Mohd~Abdullah, M.~Z. Mustaffa, F.~A. Rahim, Z.-A. Ibrahim, Y.~Yusoff,
  S.~Yussof, A.~A. Bakar, R.~Ismail, and R.~Ramli, ``Smart grid digital
  forensics investigation framework,'' in \emph{2020 8th International
  Conference on Information Technology and Multimedia (ICIMU)}, 2020, pp.
  200--205.

\bibitem{9849671}
M.~Girdhar, Y.~You, T.-J. Song, S.~Ghosh, and J.~Hong, ``Post-accident
  cyberattack event analysis for connected and automated vehicles,'' \emph{IEEE
  Access}, vol.~10, pp. 83\,176--83\,194, 2022.

\end{thebibliography}

\end{document}